\documentclass[fleqn,usenatbib]{mnras}

\usepackage{newtxtext,newtxmath}

\usepackage[T1]{fontenc}
\usepackage{ae,aecompl}

\usepackage{graphicx}	
\usepackage{amsmath}	
\usepackage{amssymb}	

\title{Planet formation and disk mass dependence in a pebble-driven scenario for low mass stars}

\author[S. Dash et al.]{
Spandan Dash$^{1}$\thanks{E-mail: dashspandan@gmail.com}
and Yamila Miguel$^{1,2}$
\\
$^{1}$Leiden Observatory, Niels Bohrweg 2, 2333 CA Leiden, The Netherlands\\
$^{2}$SRON Netherlands Institute for Space Research, Sorbonnelaan 2, 3584 CA, Utrecht, The Netherlands\\
}

\date{Accepted XXX. Received YYY; in original form ZZZ}

\pubyear{2019}

\begin{document}
\label{firstpage}
\pagerange{\pageref{firstpage}--\pageref{lastpage}}
\maketitle

\begin{abstract}
Measured disk masses seem to be too low to form the observed population of  planetary systems. In this context, we develop a population synthesis code in the pebble accretion scenario, to analyse the disk mass dependence on planet formation around low mass stars. We base our model on the analytical sequential model presented in \citet{ormel2017formation} and analyse the populations resulting from varying initial disk mass distributions. Starting out with seeds the mass of Ceres near the ice-line formed by streaming instability, we grow the planets using the Pebble Accretion process and migrate them inwards using Type-I migration. The next planets are formed sequentially after the previous planet crosses the ice-line. We explore different initial distributions of disk masses to show the dependence of this parameter with the final planetary population. Our results show that compact close-in resonant systems can be pretty common around M-dwarfs between 0.09-0.2 $M_{\odot}$ only when the disks considered are more massive than
what is being observed by sub-mm disk surveys. The minimum disk mass to form a Mars-like planet is found to be about $2 \times 10^{-3}$ $M_{\odot}$. Small variation in the disk mass distribution also manifest in the simulated planet distribution. The paradox of disk masses might be caused by an underestimation of the disk masses in observations, by a rapid depletion of mass in disks by planets growing within a million years or by deficiencies in our current planet formation picture.
\end{abstract}

\begin{keywords}
planets and satellites: formation -- planets and satellites: general
\end{keywords}



\section{Introduction}
One fascinating example of a multiple planet system is the TRAPPIST-1 system which is a compact resonant locked system of 7 exoplanets around a low mass star \citep{gillon2017seven}. Apart from the number and nature of exoplanets discovered in this system, the fact that it is a M dwarf system is even more significant. M dwarfs are the most common and longest lived of all low mass stars in the galaxy. The vast number of such stars makes them attractive options for exoplanet surveys. 
\\
\\
Since the discovery of TRAPPIST-1, exoplanet surveys have uncovered some more of these planetary systems with Earth mass planets e.g. Proxima Cen b \citep{anglada2016terrestrial} and the recently discovered Teegarden's star system \citep{zechmeister2019carmenes} and the task of actually explaining the formation of such systems is now gaining steam. Important constraints for any planet formation hypotheses around low mass stars are: (1) Explaining the formation and ubiquity of Earth mass exoplanets around M-dwarfs, as well as (2) The locked in resonant architecture of all these planetary systems. 
\\
\\
The properties of the protoplanetary disk (like its mass) and their correlation with the host star mass are important parameters that influence the planet formation process \citep{alibert2017formation}. With sub-mm surveys it has become possible to probe the disk mid-plane and estimate the dust masses in disks. However, total disk masses still remain uncertain with the scaling factor between dust mass and disk mass (the gas to dust mass ratio) still not being able to be determined precisely. Direct estimation of gas masses from molecular lines provide evidence that this scaling factor may not be universal and can vary between a factor of 10 to 1000 \citep{anglada2016terrestrial}. Moreover, disk surveys around low mass stars have only been able to characterize disks as young as 1 million years \citep{manara2018protoplanetary}.
\\
\begin{figure}
    \centering
    \includegraphics[width=\columnwidth]{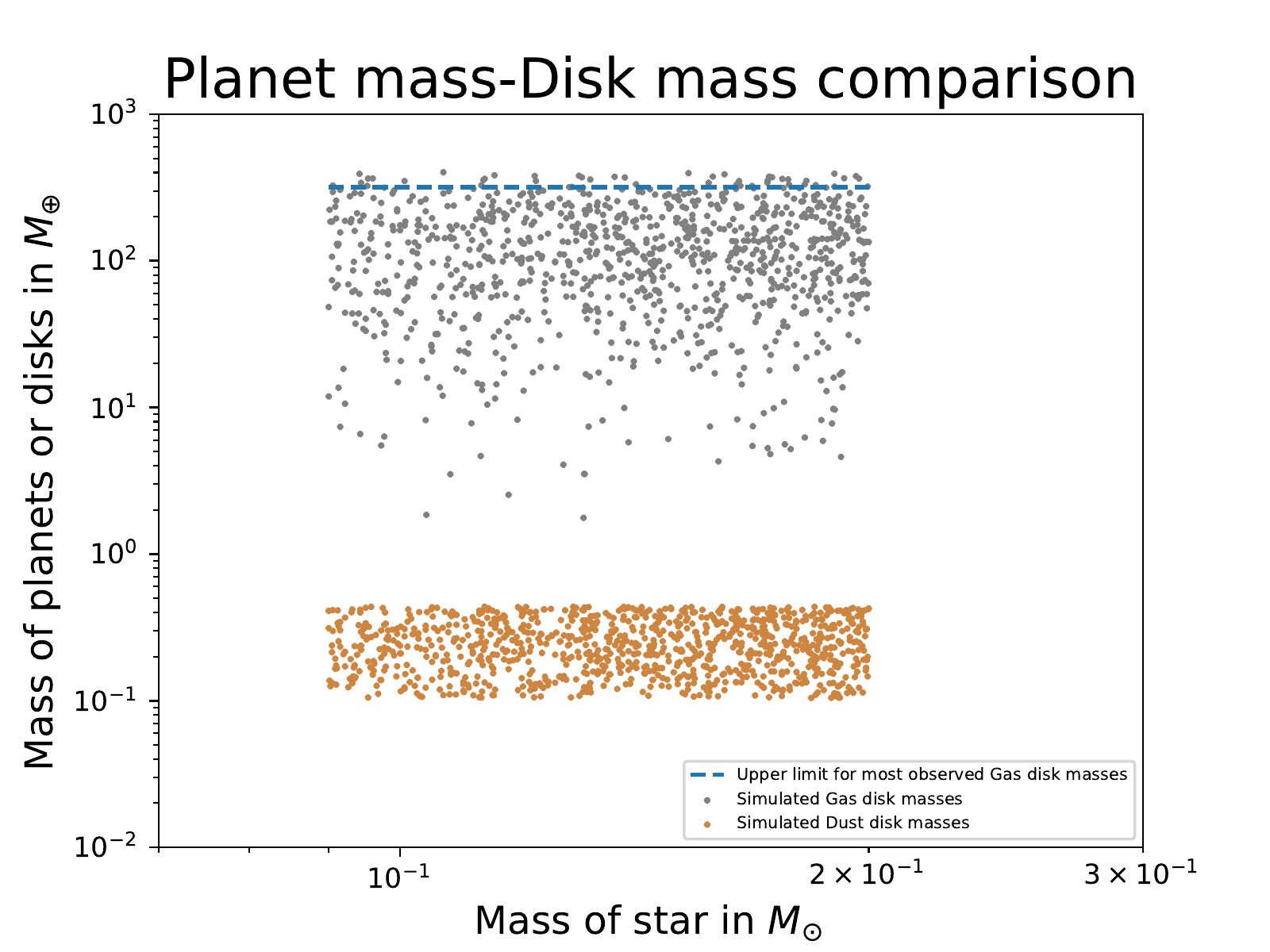}
    \includegraphics[width=\columnwidth]{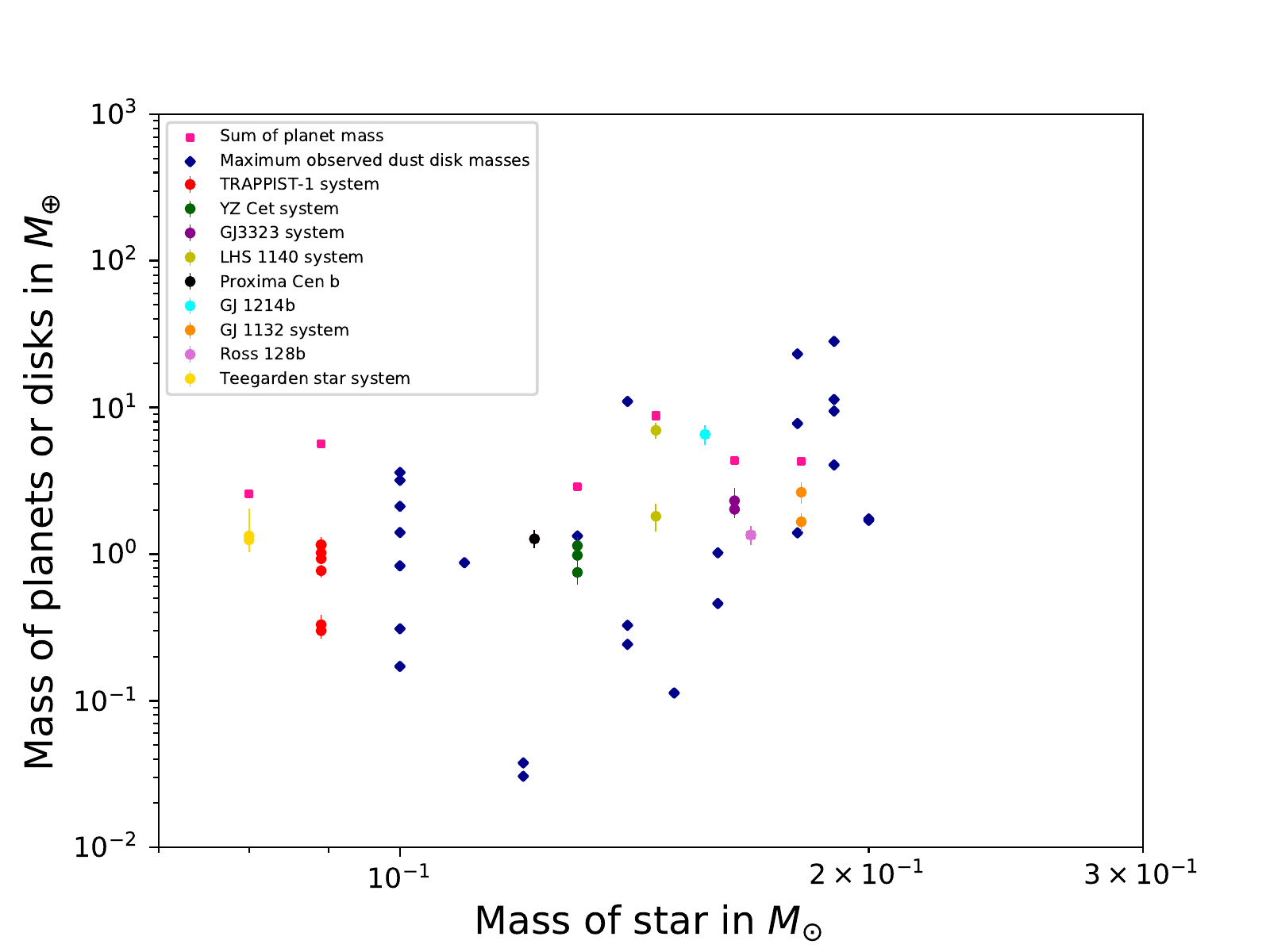}
    \caption{Comparison between (top) observed gas disk masses and 1000 simulated dust and gas disk masses around low mass M dwarf stars and (bottom) observed dust disk masses, planet masses and the upper limit of masses in these planetary systems. The dust and gas disk masses are simulated using the dust mass to star mass relation for \emph{Lupus} disks with a variable gas to dust ratio (See Section 2.3). Planet parameters are same as in Figure \ref{fig:2}. Data for observed dust and disks is from \citet{ansdell2016alma}.}
    \label{fig:i}
\end{figure}
\\
From Figure \ref{fig:i}, a majority of observed dust disk masses (for disk samples greater than a million years old) around low mass M dwarfs are either comparable to the sum of all planets in a system or not massive enough. It is also seen that while overall simulated disk masses (comparable to gas disk masses) around low mass M-dwarfs are larger than the current observed exoplanet masses, the simulated dust disk masses (from the dust mass to star mass relation for \emph{Lupus disks}, see Section 2.3) are simply not enough to form the observed planets. Either way, there seems to be a problem with explaining the presence of currently observed exoplanets around low mass M dwarfs with currently observed disk masses. \citet{manara2018protoplanetary} had also come to the same conclusion but for a wider range of stellar masses. This brings home the point of determining a minimum disk mass required to allow planetary systems similar to currently observed exoplanetary systems around low mass stars to form as well as to refine our planet formation models.
\\
\\
One thing that could explain this underestimation of disk masses is dust growth or start of planet formation within disks within 1 million years \citep{manara2018protoplanetary}. While \citet{testi2014dust} had shown that the upper limit for planet formation to start is in the 3-5 million years range, \citet{miotello2014grain} and \citet{harsono2018evidence} showed that dust growth to at least mm-cm levels has already occurred in the envelopes and disk like structures and in the inner few AU of several young YSOs (Young Stellar Objects). This shows that the start of planet formation is well underway even in the embedded stages of such YSOs. The presence of mm-cm sized dust within a million years provides impetus to the \emph{Pebble Accretion} mechanism which can quickly grow a planetesimal to a planet sized object within a million years with a mm-cm pebble reservoir already available. Keeping that in view, \emph{Pebble Accretion} has started to be increasingly used for planet formation models for growing low mass planets or giant planet cores within a million years \citep{coleman2019pebbles, brugger2020pebbles, ormel2017formation, liu2020pebbledriven}. In addition, pebble flux regulated planetesimal formation has also been proposed by \citet{lenz2019planetesimal} which has been extended by \citet{voelkel2020impact} to show that a sub-million year 100 km planetesimal formation by such a mechanism can form giant planets in the inner disk. A pebble-planetesimal hybrid accretion mechanism has also been used by \citet{alibert2018formation} to explain formation of Jupiter. All of these influence our choice to use pebble accretion as the planet formation model for our population synthesis study.
\\
\\
Several population synthesis models have now tried to understand the planet formation process in disks around low mass stars by trying to match the observed exoplanet distribution with a simulated distribution of planets. Some use the \emph{Planetesimal Accretion} model \citep{miguel2020diverse, coleman2019pebbles, alibert2017formation} and others use the more efficient \emph{Pebble Accretion} model \citep{coleman2019pebbles, liu2020pebbledriven, brugger2020pebbles}. Even the efficient \emph{Pebble Accretion} model would require dust masses larger than what is being currently observed by at least an order of magnitude since only some fraction of the total pebbles will be used for planet growth \citep{manara2018protoplanetary}. While \citet{coleman2019pebbles} showed that smaller but massive disks (10 AU with mass of 2.7-8\% of star mass) around low mass stars can explain observed planets around low mass stars, \citet{miguel2020diverse} recently concluded that a population synthesis model based on \emph{Planetesimal Accretion} alone is insufficient to explain those exoplanets if extended disks with a wide range of masses are considered (100 AU with masses between $10^{-4}$ to $10^{-2} M_{\odot}$). \citet{brugger2020pebbles} showed that comparing pebble and km sized planetesimal based accretion mechanisms with similar initial conditions produce different results with the pebble model favouring formation of Super-Earths while the planetesimal model favouring Gas Giants. Most of these models use the host star mass as a proxy and consider very massive disks from the outset. A comparative analysis of planet formation based on the initial disk mass distribution along with comparison to already  observed exoplanets hasn't been done yet. Figure \ref{fig:i} and \citet{manara2018protoplanetary}'s conclusion motivates our study where we now use the disk mass distribution as a proxy for the simulated exoplanet distribution and compare them with the observed exoplanet distribution.
\\
\\
We use the model proposed by \citet{ormel2017formation} which gave an outline for a sequential method of forming Earth mass planets in disks around low mass stars by utilizing \emph{Pebble Accretion}. The seeds for planet growth are assumed to be generated due to \emph{Streaming Instabilities} \citep{youdin2005streaming} close to the water ice line. These seeds subsequently grow by pebble accretion to Earth masses while migrating inwards by \emph{Type-I migration} into their current orbits. This sequential formation mechanism also allows these exoplanets to be trapped in resonances easily. Subsequently, a single-stage disk dispersal mechanism leaves us with a stable close-in system of exoplanets.
\\
\\
While \citet{ormel2017formation} was mostly concerned with explaining the TRAPPIST-1 system specifically by simple analytical methods, we now expand it to construct a population synthesis model in order to form and explain the entire population of exoplanets around low mass stars. We describe the construction of this model in Section 2, look at the simulated results in Section 3 and then analyze these results to comment on the problem with disk masses and compare them with results from other population synthesis models in Section 4.

\section{Methods}
\subsection{Disk Model and Temperature Structure}
The gas surface density profile of the disk ($\Sigma_{g}$) is assumed to be a simple power law disk profile with:
\begin{equation}
    \Sigma_g = \frac{(2-p)M_{disk}}{2 \pi a_{out}^{2}}\bigg{(}\frac{a}{a_{out}}\bigg{)}^{-p}.
\end{equation}
Here, $M_{disk}$ is the total mass of the disk, $a_{out}$ is the disk outer radius and $a$ is the semi-major axis. While this profile can be used for any $0< p < 2$, we adopt a value of $p=1$.
\\
\\
The disk temperature profile at a semi-major axis $a$ (in AU) is:
\begin{equation}
    T(a) = 180K\frac{M_{\star}}{0.08 M_{\odot}}\bigg{(}\frac{h}{0.03}\bigg{)}^{2}\bigg{(}\frac{a}{0.1}\bigg{)}^{-1}.
\end{equation}
Here, $M_{\star}$ is the mass of the star and $h$ is the disk aspect ratio (a measure of the disk vertical profile in comparison to its distance from the star) which is assumed to be $0.05\times(a/1AU)^{1/4}$ \citep{ida2004toward}. Equation 2 can be modified, by using the equation for $h$ and the ice-point temperature to be at 180 K, to give (in AU) the distance to the ice-line at:
\begin{equation}
    a_{ice-line} = 0.077\bigg{(}\frac{M_{\star}}{0.08M_{\odot}}\bigg{)}^{2}.
\end{equation}
\\
\\
The magnetospheric cavity radius is taken to be the point at which the inner disk is truncated. Using a value of stellar radius as 0.5 $R_{\odot}$ and stellar magnetic field of 180 G, the inner disk radius is (in AU) \citep{ormel2017formation}:
\begin{equation}
    a_{inneredge, disk} = 0.0102 \bigg{(}\frac{0.08M_{\oplus}}{M_{\star}}\bigg{)}^{1/7}\bigg{(}\frac{10^{-10}M_{\odot}yr^{-1}}{\dot{M}_{g}}\bigg{)}^{2/7}.
\end{equation}
Here, $\dot{M}_{g}$ is the gas accretion rate from the viscous accretion disk into the star.

\subsection{Planet Formation Model}
\subsubsection{Growing the Planets}
Since our model is a population synthesis extension to the analytical model proposed by \citet{ormel2017formation}, we refer to that paper for the details of the parameters in the model. However, we explain these briefly here for the sake of completeness. The planet growth equation using pebble accretion can be written as:
\begin{equation}
    \dot{M}_{pl} = \epsilon F_{p/g} \dot{M}_{g}.
\end{equation}
Here, $\dot{M}_{pl}$ is the differential mass growth with time. $\epsilon$ is the efficiency of mass growth from a radially drifting pebble front. $F_{p/g}$ is the pebble to gas mass flux generated due to the same radially drifting pebble front. This front is obtained due to growth from micron sized particles by collisions. $\dot{M}_{g}$ is assumed to be $10^{-10}M_{\odot}$yr$^{-1}$ \textcolor{blue}{\citep{ormel2017formation}}. $F_{p/g}$ is defined from the disk properties as:
\begin{equation}
    F_{p/g} = \frac{2M_{disk}Z_{0}^{5/3}}{3\dot{M}_{g}a_{out}\kappa^{2/3}}\bigg{(}\frac{GM_{\star}}{t}\bigg{)}^{1/3}.
\end{equation}
Here, $Z_{0}$ is the metallicity of the disk and is assumed to be 0.02 and $\kappa$ is the number of e-foldings (growth by a factor of e at each step) needed to reach pebble size and is assumed to be 10 \textcolor{blue}{\citep{ormel2017formation}}.
\\
\\
The seeds for pebble accretion are assumed to be formed from streaming instability near the water ice-line where the mid-plane pebble to gas density ratio can exceed 1 under specific parameters for viscosity ($\alpha \geq 10^{-3}$) of the disk \citep{schoonenberg2017planetesimal,ormel2017formation}. We also assume an ice-line width of 0.02 AU in which the seeds can be produced which is motivated by the width over which the pebble to gas density ratio remains above 1 as shown by \citet{ormel2017formation}.
\\
\\
As soon as seeds of about $10^{-4} M_{\oplus}$ (equivalent to mass of Ceres) are available, pebble accretion is assumed to start \citep{bitsch2015growth}. The first phase of mass growth is within the ice-line width we assumed above and most pebbles accreted onto the seed are icy pebbles. With the value of $\alpha$ we have ($\alpha \geq 10^{-3}$), the mass growth mechanism in this case is planar (2D) \citep{ormel2017formation}) and Equation 5 becomes:
\begin{equation}
     \dot{M}_{pl,2D} = \epsilon_{2D} F_{p/g} \dot{M}_{g}.
\end{equation}
Here, the value of $\epsilon_{2D} = 184.6 \times q_{pl}^{2/3}$. $q_{pl}$ is defined as $M_{pl}/M_{\star}$ where $M_{pl}$ is the mass of the planet.
\subsubsection{Migrating the planets}
The growing core is then subject to Type I migration, first towards the ice-line inner edge and then further till it reaches the inner edge of the disk. The equation for Type-I migration is \citep{lambrechts2014forming}:
\begin{equation}
    \frac{da}{dt} = -2.8 q_{pl}c_{mig} \frac{\Sigma_{g}a^{2}}{M_{\star} h^{2}}v_{K}.
\end{equation}
Here, $v_{K}$ is the Keplerian velocity at a particular $a$, $c_{mig}$ is the reduction factor used as a measure of uncertainties and non-linear effects affecting migration. In this paper, we use two values of $c_{mig}$ as 1 and 0.1 and we call them standard migration and reduced rate migration models respectively in a similar vein to how it was done in  \citet{miguel2020diverse} and \citet{ida2004toward}.
\\
\\
The planetary core migrates until it reaches the inner edge of the ice-line and then migrates further inwards into a drier part of the disk. Now the pebbles being accreted are smaller silicate pebbles and the mass growth is more inefficient. Equation 5 now becomes:
\begin{equation}
    \dot{M}_{pl,3D} = \epsilon_{3D} F_{s/g} \dot{M}_{g}.
\end{equation}
Here, $\epsilon_{3D} = 0.07 \times (q_{pl}/10^{-5})$ and $F_{s/g}$ is the silicate to gas mass flux. \citet{schoonenberg2017planetesimal} assume that at most 50\% by mass of the pebble beyond the ice-line is composed of ice. Interior of the ice-line, this ice evaporates and hence we assume a reduced metallicity of 0.5$Z_{0}$ in Equation 6 to account for the missing mass from ice and calculate $F_{s/g}$.
\subsubsection{Stopping of planet growth}
The planet migration remains Type I till it ultimately reaches the inner edge of the disk. However, the mass growth stops at the \emph{Isolation Mass} at which point the planet starts forming a gap in the disk and is hence cut off from its pebble supply. This mass is determined as \citep{ataiee2018much}:
\begin{equation}
    M_{iso,pl} \approx h^{3}\sqrt{37.3\alpha+0.01} \times \bigg{[}1+0.2\bigg{(}\frac{\sqrt{\alpha}}{h}\sqrt{\frac{1}{St^{2}}+4}\bigg{)}^{0.7}\bigg{]}\times M_{\star}.
\end{equation}
Since streaming instabilities are only possible within the ice-line for $\alpha$ > $10^{-3}$, we assume the lowest value of $10^{-3}$ for this calculation. $St$ is the Stokes Parameter and for efficient pebble accretion has the lowest value of 0.05 \citep{ataiee2018much, bitsch2018pebble} which we assume here. The pebble isolation mass ensures that not only do the planets stop accreting pebbles but also that the inward drift of pebbles is also stopped \citep{bitsch2018pebble}. 
\\
\\
Planet mass growth is also stopped when the radial pebble flux stops. This happens when the outward moving pebble front reaches the outer disk edge. The time when this happens is \citep{ormel2017formation}:
\begin{equation}
    t_{end} = \frac{\kappa}{Z_{0}}\bigg{(}\frac{a_{out}^{3}}{GM_{\star}}\bigg{)}^{0.5}.
\end{equation}
\subsubsection{Slow Gas Accretion regime}
After the planet reaches an isolation mass, the mass accretion moves into a period of slow gas envelope accretion and compression which is modeled as \citep{bitsch2015growth} (in terms of $M_{\oplus}$ masses/million years):
\begin{multline}
    \frac{dM_{g}}{dt} = 0.00175f^{-2}\bigg{(}\frac{\kappa_{env}}{1 cm^{2}/g}\bigg{)}^{-1}\bigg{(}\frac{\rho_{c}}{5.5g/cm^{3}}\bigg{)}^{-1/6}\bigg{(}\frac{M_{c}}{M_{\oplus}}\bigg{)}^{11/3} \\
    \bigg{(}\frac{M_{g}}{0.1 M_{\oplus}}\bigg{)}^{-1}\bigg{(}\frac{T}{81 K}\bigg{)}^{-0.5}.
\end{multline}
This can be further simplified a bit by assuming $\kappa_{env}$ and $\rho_{c}$ to be 1 cm$^{2}$/g and 5.5g/cm$^{3}$ respectively, f is 0.2, $M_{c}$ is the isolation mass and initial $M_{g}$ is 0.1 times the isolation mass \citep{bitsch2015growth}. Equation 12 then becomes (in terms of $M_{\oplus}$ masses/year):
\begin{multline}
    \frac{dM_{g}}{dt} = 1.771 \times 10^{-10} \bigg{(}\frac{M_{iso,pl}}{M_{\oplus}}\bigg{)}^{11/3}\bigg{(}M_{gas}\bigg{)}^{-1}
    \\
    \bigg{(}\frac{M_{\star}}{0.08M_{\odot}}\bigg{)}^{-1/2}\bigg{(}\frac{a}{1AU}\bigg{)}^{1/4}.
\end{multline}
This slow accretion continues till the planet and envelope mass together reaches a critical mass of about 2$M_{iso,pl}$ \citep{bitsch2015growth} after which there is rapid gas accretion onto the envelope. However, we find that none of our planets manage to reach this limit. The amount of gas accreted is also very meagre and can easily be stripped off when the disk eventually dissipates. This means that we can safely neglect this contribution for our simulations. This is similar to what was found in \citet{coleman2019pebbles} who used a more precise self-similar gas accretion model.

\subsubsection{Formation of a resonant convoy of planets}
We assume that when the first planet reaches the inner edge of the disk, it stops there and continues growing at the same efficiency it had before. In the absence of any concrete hypothesis of events happening in such a case, this simplifying assumption is made just for the sake of calculation and can be easily modified by introducing a fudge factor for the efficiency of either pebble or gas accretion.
\\
\\
We assume that the second planet mass growth is triggered as soon as the first planet crosses the ice-line. All steps as outlined in the previous subsections are followed again until the planet reaches a minimum separation distance from the first planet and stops migrating or the gas disk is dispersed which also stops migration. 
\\
\\
We assume that two planet interactions here are mostly independent of an external influence. This is similar to the assumption made in \citet{sasaki2010origin} (for a Jovian moon resonant system) and \citet{miguel2020diverse} (for exoplanets around low mass stars formed with core accretion mechanism). In such a case, orbits of two planets will be stable if they are separated by $K \times r_{H,m}$ where $K$ is a critical parameter for planet spacing, $r_{H,m}$ is the mutual Hill radius and is equal to $((a_{1}+a_{2})/2)\times((m_{1}+m_{2})/3M_{\star})^{1/3}$. Here, $a_{i}$ and $m_{i}$ correspond to semi major axis and mass of the ith planet. For critical parameter $K$, we base our assumption on \citet{pu2015spacing} who found that compact multi-planetary systems can be sculpted into systems with wide spacing. They found that systems where this spacing among planets is characterised by $K \leq 8.1$ can be weeded out in the nascent disk and hence orbits with spacing greater than this amount of mutual Hill Radii should be stable on a billion year timescale. On the other hand, they also found that Kepler planets seem to be tightly clustered around a value of $K = 12$. Motivated by these threshold values, we assume a minimum spacing between each pair of planets at random in between these two limits for our simulations.
\\
\\
An additional complication to keep in mind here is the possibility of a planet upstream reaching pebble \emph{Isolation mass} before the planet downstream which would then cut of pebble supply for the planet growing downstream. However, from our simulations we find that the second planet always has to migrate more slowly (See Section 2.2.6) and has to reach a higher \emph{Isolation mass} than the first planet as a result. This means that our model avoids this complication and the interference on the growth of the planet downstream by the planet upstream is reduced even though our model is a sequential formation model. Subsequently, more planets are formed and migrate to their designated orbits before the disk dissipates allowing the formation of multiple planetary systems.
\subsubsection{Gaseous disk evolution}
Observations indicate that the gas disk around stars dissipates on a scale of 1-10 million years \citep{hartmann1998accretion, bitsch2015growth}. However, the photoevaporation rate after 3 million years can substantially deplete the disk of gas \citep{alexander2014protostars}. \citet{pecaut2016star} found that disks around K/M stars can last longer than this lifetime. However, we find that the pebble flux stopping time from Equation 11 takes less than 1 million years for the stars we consider in our simulations. This means that most of the disk lifetime would then only be for the orbital evolution of already formed planets. Hence, similar to \citet{bitsch2015growth}, we take our disk lifetimes in the range of 1 to 3 million years. We note that taking a longer timescale would slightly affect the orbital positions of planets not already stuck in mean motion resonances.
\\
\\
Gaseous disk evolution will affect how our planets grow and migrate. To model this, we use a simple exponential decay function \citep{ida2004toward}:
\begin{equation}
    \Sigma_{g} = \Sigma_{g,0} e^{-\frac{t}{\tau}}.
\end{equation}
$\Sigma_{g,0}$ is the gas surface density profile at the beginning of the calculation (Equation 1), $\tau$ is the disk dissipation lifetime which can vary between 1 and 3 million years and $t$ is the current disk age. The exponential contribution term is named as the \emph{dissipation factor}. As the gas continuously depletes from the disk, the gas accretion rate into the star also slows down as:
\begin{equation}
    \dot{M}_{g} = \dot{M}_{g,0} e^{-\frac{t}{\tau}}.
\end{equation}
From Equations 6 and 5, it is easy to see that the product $F_{p/g}\dot{M}_{g}$ is independent of the dissipation factor. This means that the efficiency of accretion of planets remains independent of the disk age. Equations 8 and 14 makes it clear that the dissipation factor would result in slowing of the planet orbital migration rate as the disk grows older. Change in other parameters of the model is mentioned in Section 2.3.
\subsection{Initial model parameters}
\begin{figure}
    \centering
    \includegraphics[width = \columnwidth]{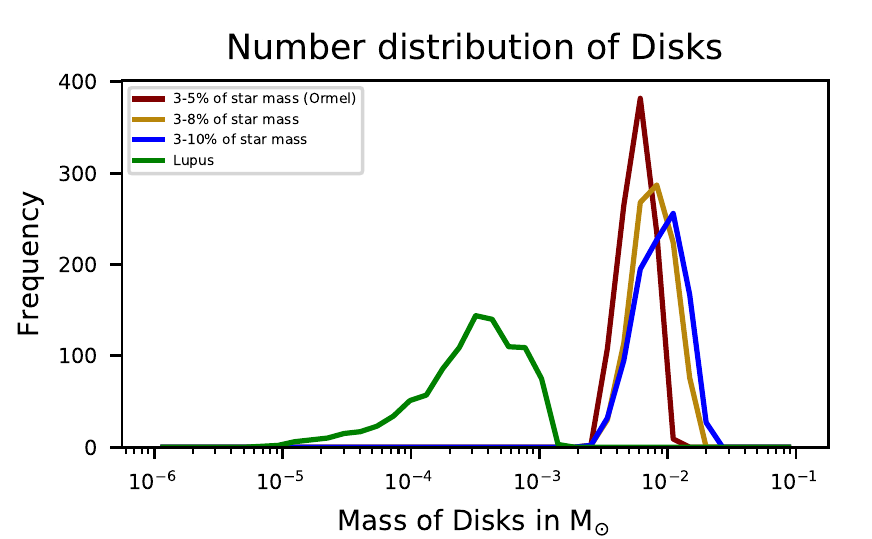}
    \caption{Comparison of all simulated disk mass distributions.}
    \label{fig:0}
\end{figure}

The model we use is made from scratch in Python 2.7 and is available online\footnote{https://github.com/dashspandan/planet}. We consider star masses from 0.09 $M_{\odot}$ to 0.2 $M_{\odot}$. We select a mass at random for each iteration from a uniform list of masses in the above mentioned range. Given the uncertainty in determining disk mass around low mass stars, we assume two possibilities. The first is an arbitrary disk mass to star mass relation of 0.03-0.05 (average valued of 0.04 used by \citet{ormel2017formation} and hence these disks are herewith called as \emph{Ormel disks}) and the other is a more precise disk dust mass to star mass relation obtained from observations in the Lupus and Taurus molecular clouds \citep{ansdell2016alma}. There is evidence that the gas to dust mass value in disks is not 100 \citep{bohlin224survey} and instead varies between 10 to 1000 \citep{ansdell2016alma}. Hence, we use a value of gas to dust mass ratio ($g/d$) from a uniform list between 10-1000 to scale this relation to a disk mass-star mass relation. The modified relation from \citet{ansdell2016alma} becomes:
\begin{equation}
log M_{disk} = 1.8 log M_{\star} + 0.9 + log(g/d) + log(3\times10^{-6}).
\end{equation}
These disks will be henceforth called as \emph{Lupus disks} for convenience sake. To incorporate more variety in our disk samples, we also consider disk masses between 3-8\% and 3-10\% star mass which are more massive than \emph{Ormel disks}. We show all the disk mass distributions in Figure \ref{fig:0}. For each iteration, we choose one disk at random from the distributions we have constructed based on our runs (Run 1 through 4, See Table \ref{tab:1}).
\\
\\
Apart from the above discussed main independent parameters, there are also several other dependant parameters which we list below:
\begin{itemize}
    \item Disk dispersal lifetime: We pick a random age from a uniform list of ages from 1 to 3 million years (Section 2.2.6).
    \item Water ice-line: We use Equation 3 for this.
    \item Disk inner radius: We use Equation 4 for this. This value will increase with time as it is inversely proportional to $\dot{M_{g}}$ (Equation 15). We assume that any inwards migrating planet will be stopped when it reaches this disk edge (similar to the assumption in \citet{coleman2016formation} for non-gap forming planets). 
    \item Disk outer radius ($a_{g}$): This value is assumed to be 200 AU \citep{ormel2017formation}, which falls in the range of observed outer radii of gas disks \citep{ansdell2018alma}.
    \item Disk metallicity ($Z_{0}$): This is taken to be 0.02 (super-solar) to ensure that sufficient rocky material is available to form multiple planets as well as to make streaming instabilities easy by increased clumping of dust particles \citep{johansen2009particle}.
    \item Minimum spacing between planets: The spacing between each planet pair is chosen at random from a uniform list of separations between 8 and 12 $r_{H,m}$ (See Section 2.2.5). For our simulations, simplifying this in terms of the Hill radius of just the preceding planet would make computation easier. We assume that $m_{1} \sim m_{2}$ and $a_{1} \sim a_{2}$ which is approximately valid for close in compact systems like ours where consecutive planets are found to be quite similar in final masses.
    \item Maximum number of planets: We limit the maximum number of planets that can be formed in a planetary system to 20 due to numerical constraints and keeping in view the complexities of large N. Since all planets are formed inside the ice-line width and migrate inwards, the disk inner edge and the outer edge of the ice-line width form the innermost and outermost extent of the semi-major axis distribution for planets. We observe that all disks in our reduced rate migration model and in the standard migration model form less planets in a system than the maximum limit.
\end{itemize}

\section{Results}
\subsection{Planet evolution profiles}
As mentioned in Section 2.2.2, we consider two different values of $c_{mig}$ for our analysis i.e. 1 and 0.1 (standard model and reduced migration model respectively).
\subsubsection{With $c_{mig}=1$(standard model)}
\begin{figure}
    \centering
    \includegraphics[width = \columnwidth]{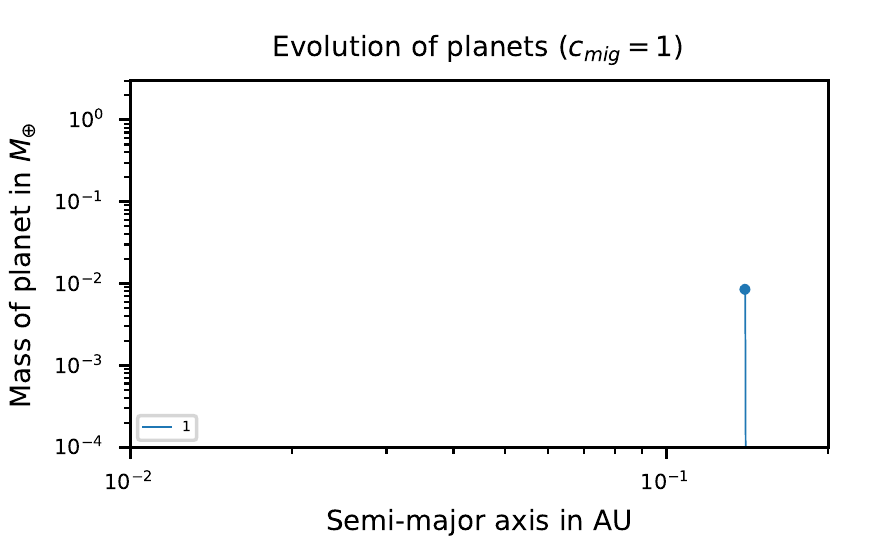}
    \includegraphics[width = \columnwidth]{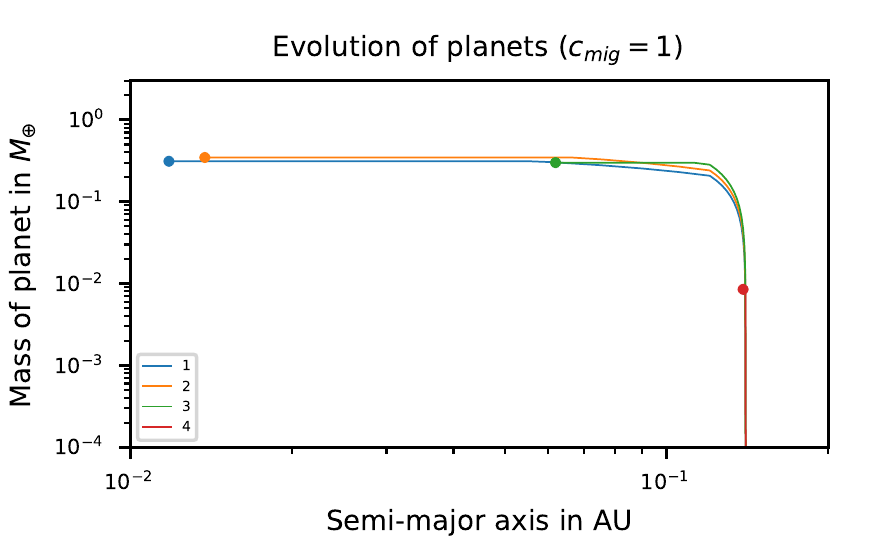}
    \includegraphics[width = \columnwidth]{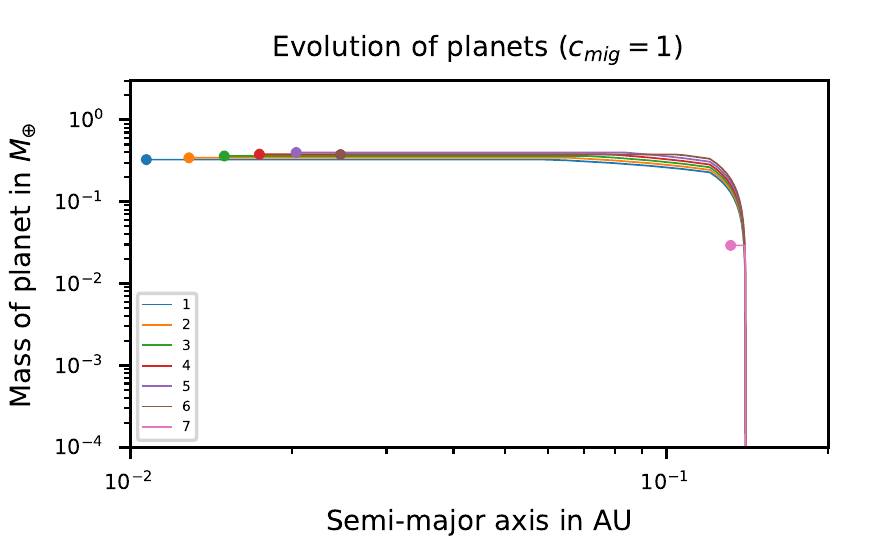}
    \caption{The effect of increasing disk masses on planetary system evolution around a 0.1 $M_{\odot}$ star. Disk masses are: (top) \emph{Lupus} disk with $g/d = 1000$, (middle) 3\% star mass and (bottom) 5\% star mass.}
    \label{fig:1}
\end{figure}

To determine the effect of disk masses on a planetary system's evolution, we run our simulation for a 0.1 $M_{\odot}$ star and three different disk masses, evolving the system in all cases for 1 million years. Three different cases of disk masses are used and the results are shown in Figure \ref{fig:1}: (top) \emph{Lupus} disk with $g/d = 1000$, (middle) Disk mass is 3\% star mass and (bottom) disk mass is 5\% star mass.
\\
\\
In Figure \ref{fig:1} we see that the effect of increasing disk masses mostly affects the number of planets formed per system (top panel has 1, middle panel has 4 and bottom panel has 7). This is because migration rate is faster around more massive disks (from Equations 1 and 8) and hence planets can cross the ice-line fast enough to trigger the formation of subsequent planets in our model. In general, Figure \ref{fig:1} shows that migration in planets starts being significant at around the Mars mass limit. Hence, increasing disk masses also means that most planets in the system can cross this threshold in order to migrate away from the ice-line.
\\
\\
In the middle panel of Figure \ref{fig:1}, the first two planets are separated by the minimum spacing required for stable orbits for both while the other two stopped migrating before reaching that limit due to disk dispersal. The second planet is more massive than the first planet and also more massive than the third and fourth planets as the pebble supply ends (Equation 11) after it reaches its \emph{Isolation mass} limit but before the third planet reaches this limit. Regarding migration, since each succeeding planet starts evolving a bit later than its predecessor, its migration is slower (due to gas depletion which slightly reduces the gas mass in the disk; from Equations 8 and 14) while the mass accretion rate is constant (See Section 2.2.6). This allows it to accrete more mass at the same value of semi-major axis.
\\
\\
In the bottom panel of Figure \ref{fig:1}, the effect of a succeeding planet accreting more mass at similar semi-major axis is more pronounced. Since migration rate is the fastest, more planets reach their respective \emph{Isolation masses} (with the first planet quickly reaching a lower value of disk inner radius (See Equation 4)). As succeeding planets accrete more mass at similar semi-major axis, each succeeding planet manages to reach its \emph{Isolation mass} at a larger value of the semi-major axis. Hence in this system, the first 5 planets form a convoy with increasing order of mass as we move outward. The 6th and 7th planets cannot accrete enough mass as the pebble supply stops before they can reach the \emph{Isolation mass} limit (Equation 11). The 7th planet stops migrating before it can migrate to an orbit with the lowest limit for spacing from its preceding planet. Every other planet before it manages to reach that limit before disk dispersal.
\subsubsection{With $c_{mig}=0.1$(reduced rate migration model)}
\begin{figure}
    \centering
    \includegraphics[width = \columnwidth]{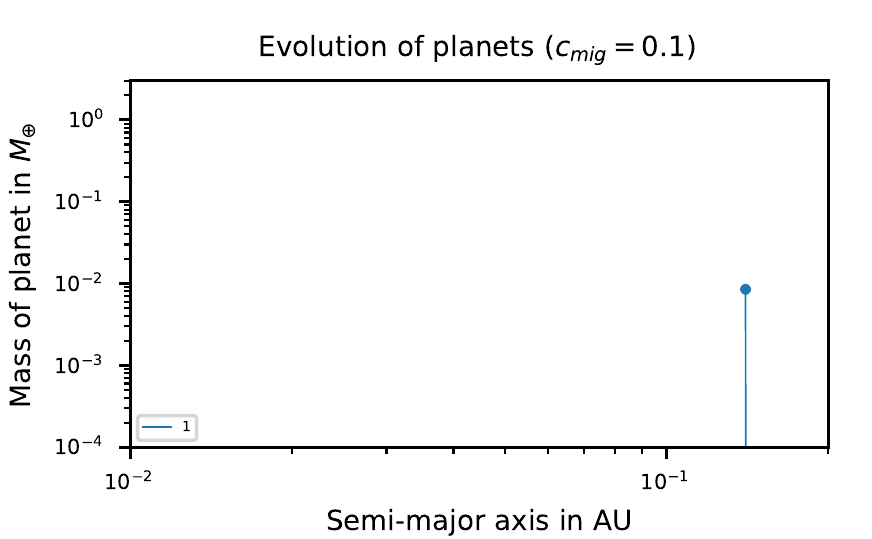}
    \includegraphics[width = \columnwidth]{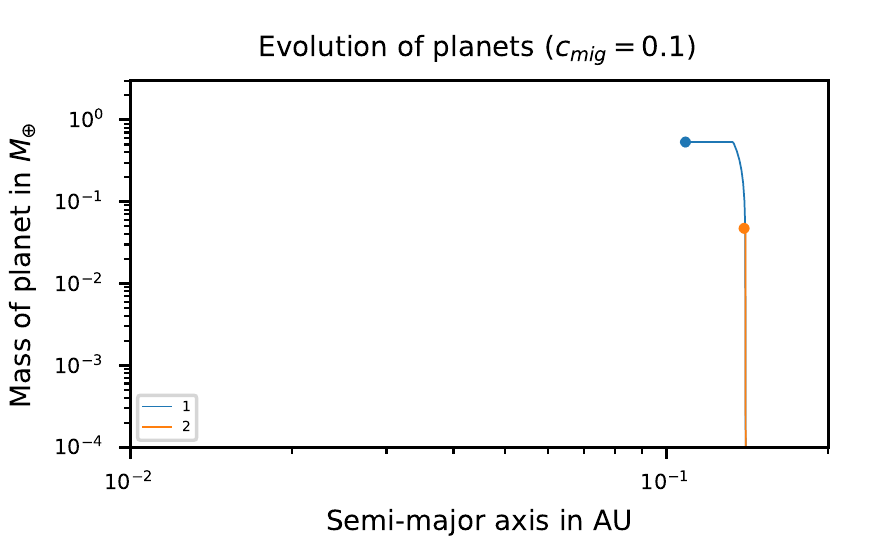}
    \includegraphics[width = \columnwidth]{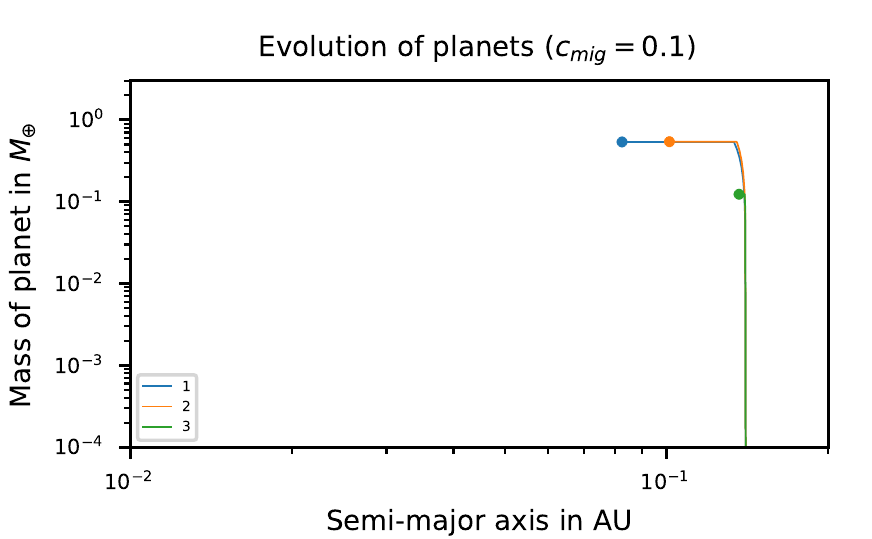}
    \caption{The effect of increasing disk masses on planetary system evolution around a 0.1 $M_{\odot}$ star. Disk masses are: (top) \emph{Lupus} disk with $g/d = 1000$, (middle) 3\% star mass and (bottom) 5\% star mass. The rate of migration has been reduced by a factor of 10 for these simulations.}
    \label{fig:1b}
\end{figure}

We now reduce the migration rate by a factor of 10 for the same set of three simulations done above to see the effect of a longer migration timescale for same disk masses.
\\
\\
From Figure \ref{fig:1b} we see that the effect of delaying the migration rate by a factor of 10 leads to a reduced number of planets for all simulation except \emph{Lupus disks} (top has 1, middle has 2 and bottom has 3) for each system for similar disk masses. This is expected for heavier disks as slower migration means that a lesser number of planets can cross the ice-line and trigger subsequent planet formation. For \emph{Lupus disks}, the planet grows to similar mass as the accretion rate remains unchanged due to slower migration but the planets can't grow to Mars mass limit to cross the ice-line in both cases. The slower migration means that most planets manage to reach their \emph{Isolation masses} inside the ice-line width at larger values of semi-major axis. This means that the most massive planets formed in each system is more massive than its counterpart for faster migration.
\subsection{Population synthesis with varying disk masses}
\subsubsection{$c_{mig} = 1$(standard model)}
\begin{figure*}
    \centering
    \includegraphics[width=\columnwidth]{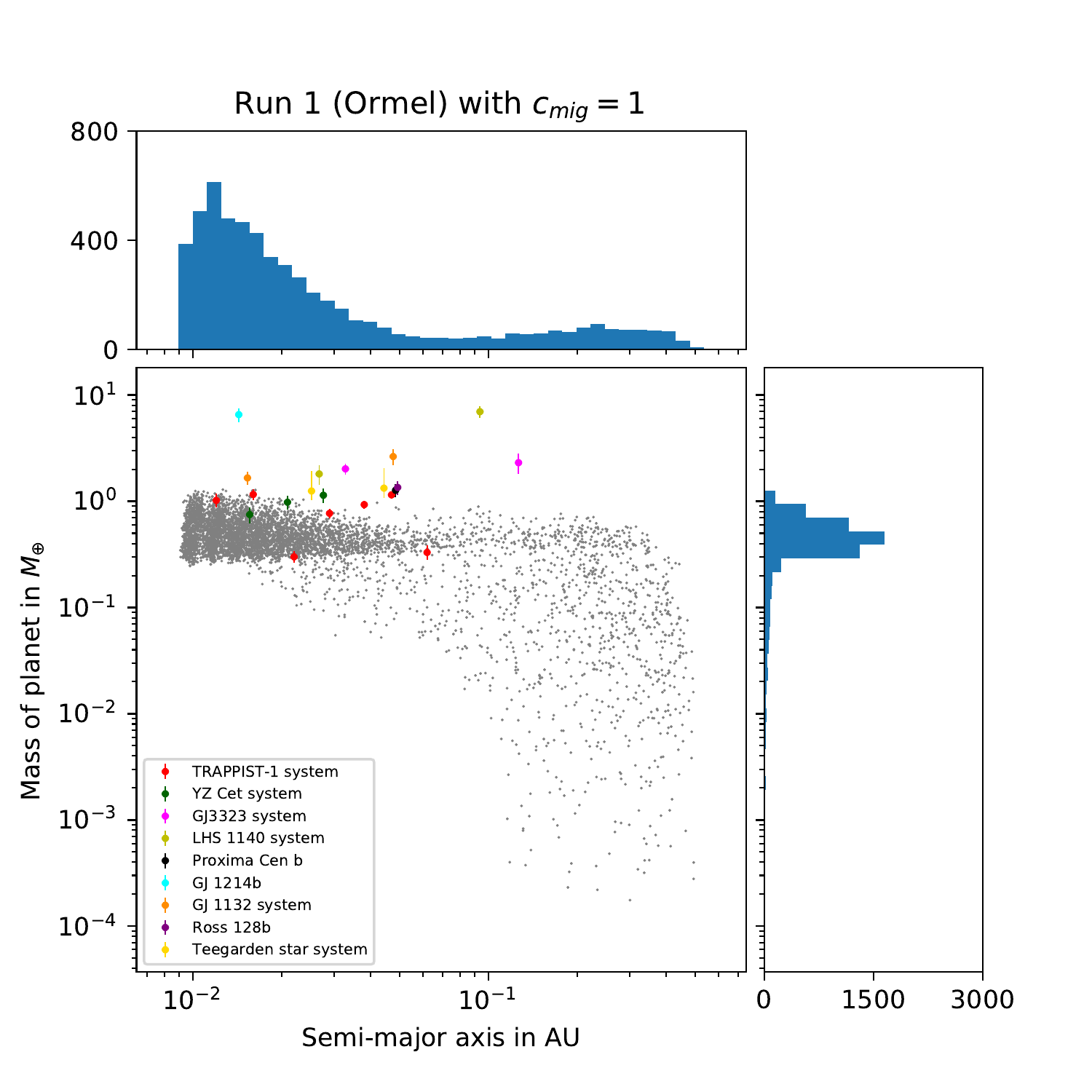}
    \includegraphics[width=\columnwidth]{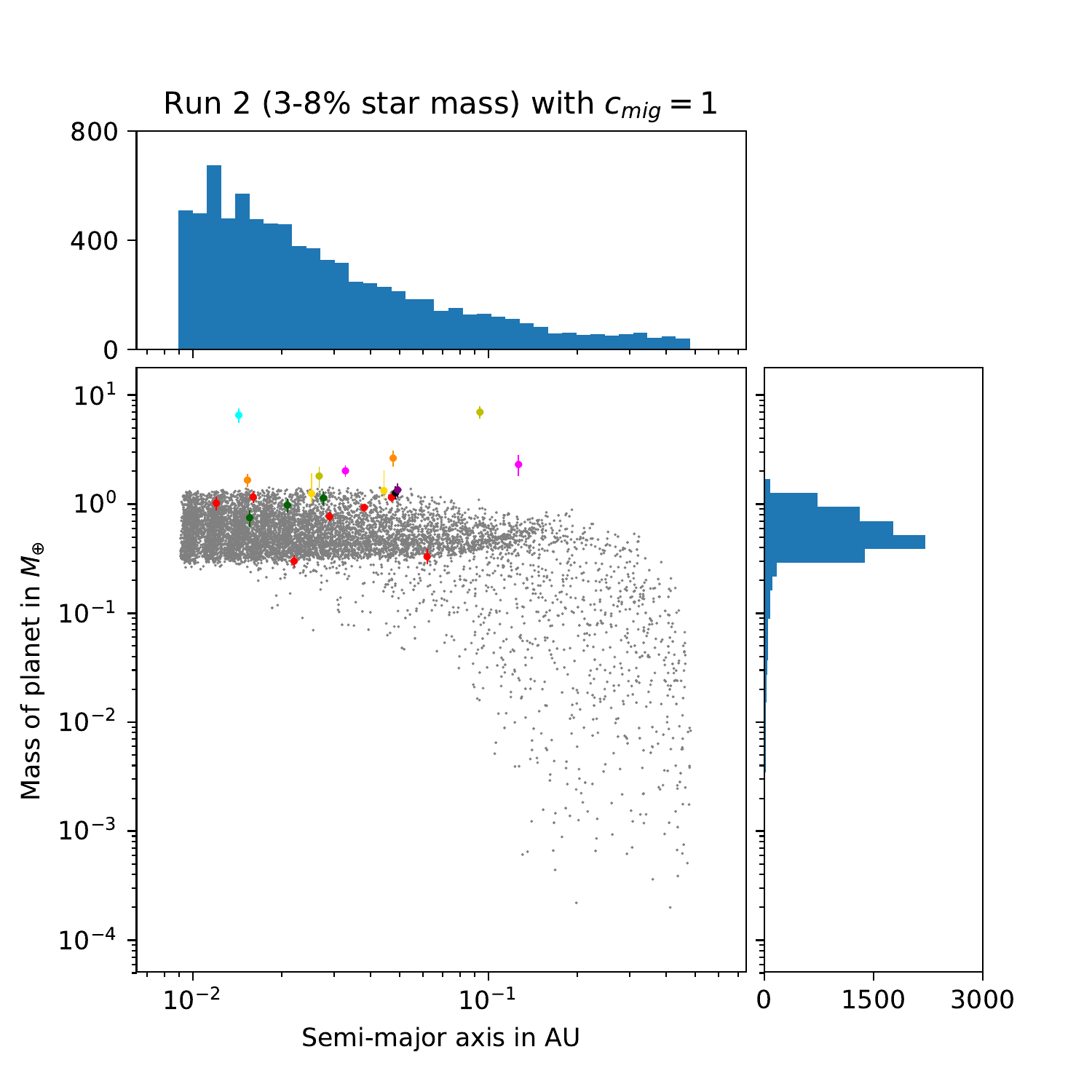}
    \includegraphics[width=\columnwidth]{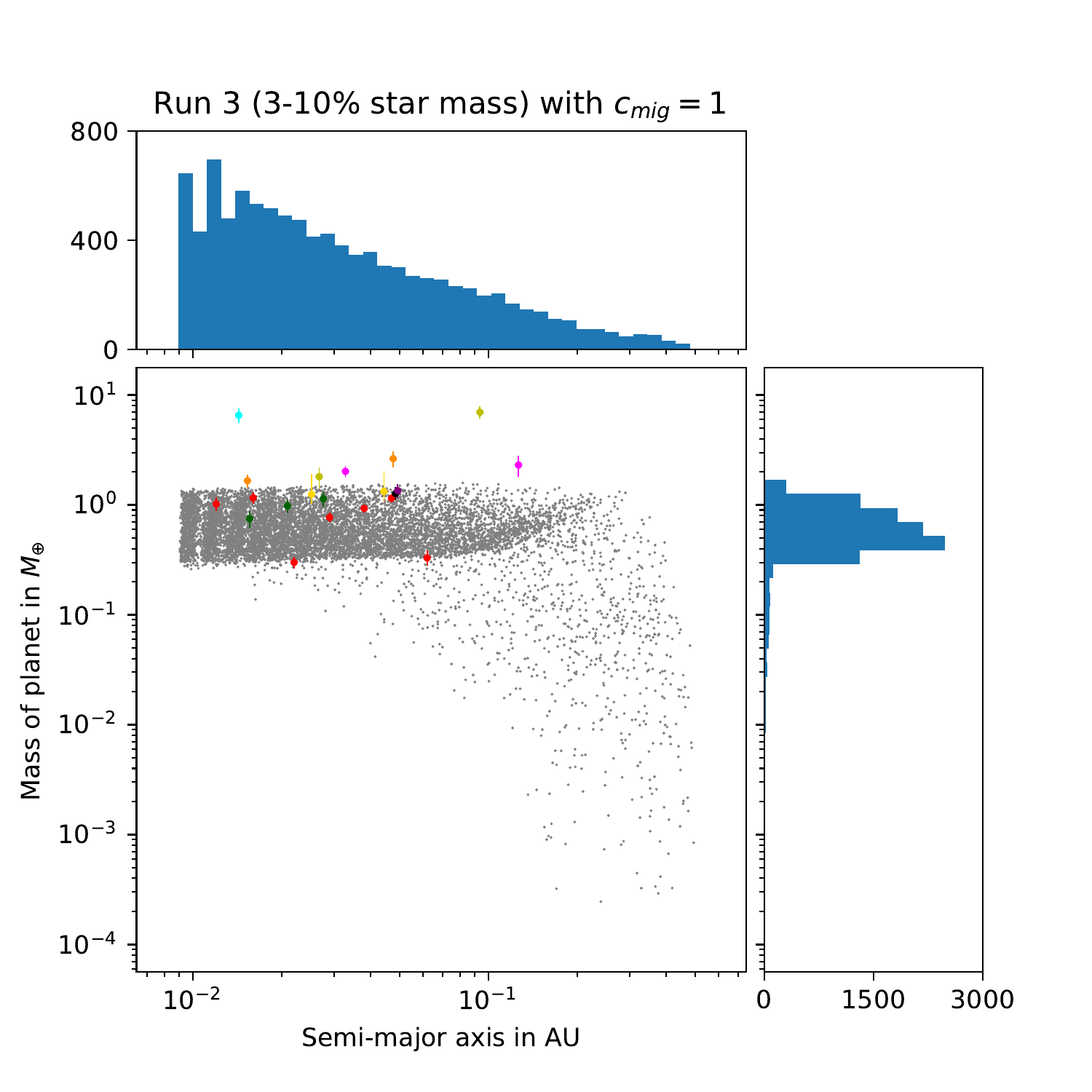}
    \includegraphics[width=\columnwidth]{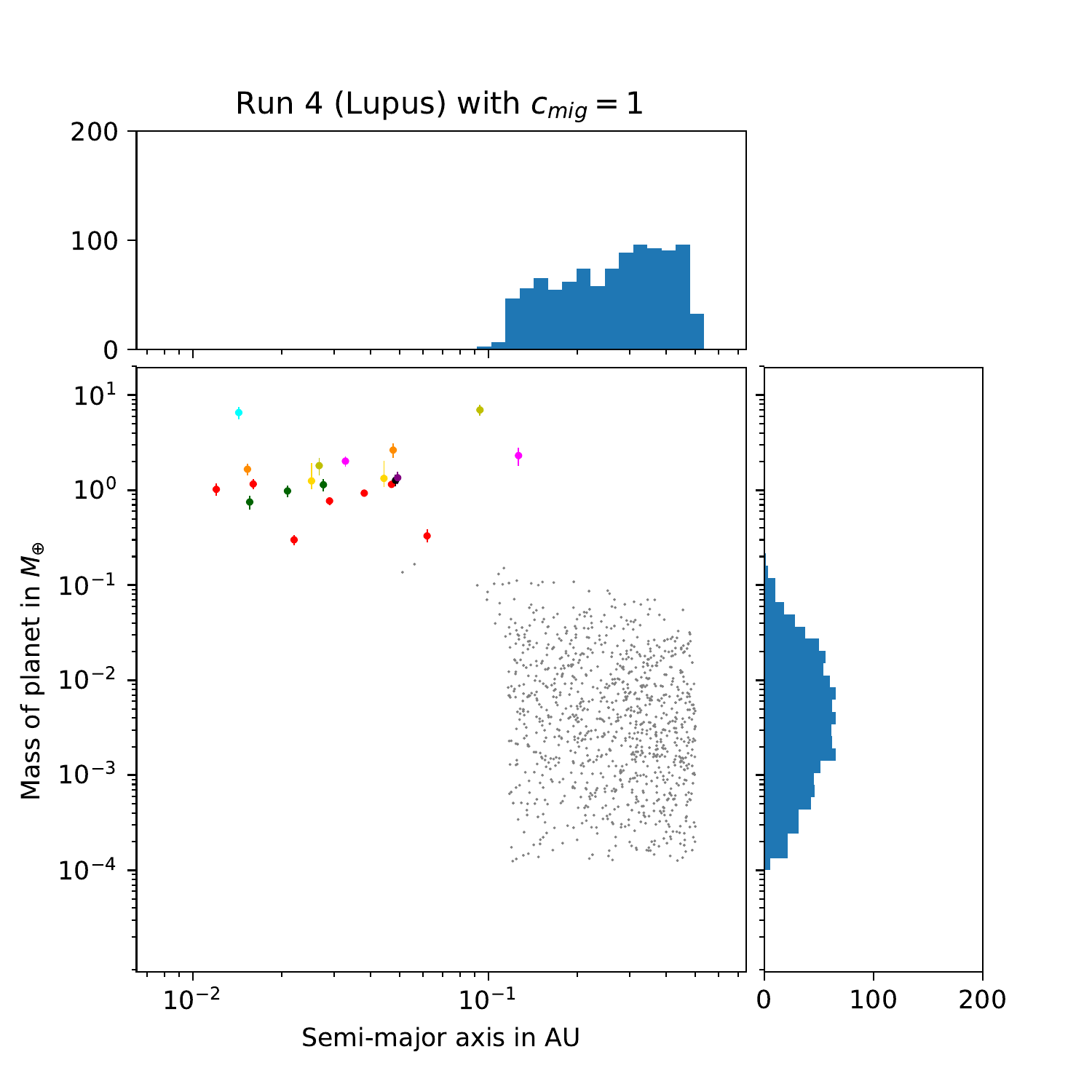}
    \caption{Synthetic population of planets for (top left) \emph{Ormel} Disks (3-5\% star mass), (top right) a bit more heavier disks (3-8\% star mass), (bottom left) a lot more massive disks (3-10\% star mass) and (bottom right) \emph{Lupus} disks. The parameters of planets in each observed system are taken from: TRAPPIST-1 \citep{grimm2018nature}, YZ Cet system \citep{astudillo2017harps1}, GJ 3323 system \citep{astudillo2017harps2}, LHS 1140 system \citep{ment2018second}, Proxima Cen b \citep{anglada2016terrestrial}, GJ 1214b \citep{charbonneau2009super}, GJ 1132 system \citep{bonfils2018radial}, Ross 128b \citep{bonfils2018temperate} and Teegraden's star system \citep{zechmeister2019carmenes}.}
    \label{fig:2}
\end{figure*}
\begin{figure}
    \centering
    \includegraphics[width=\columnwidth]{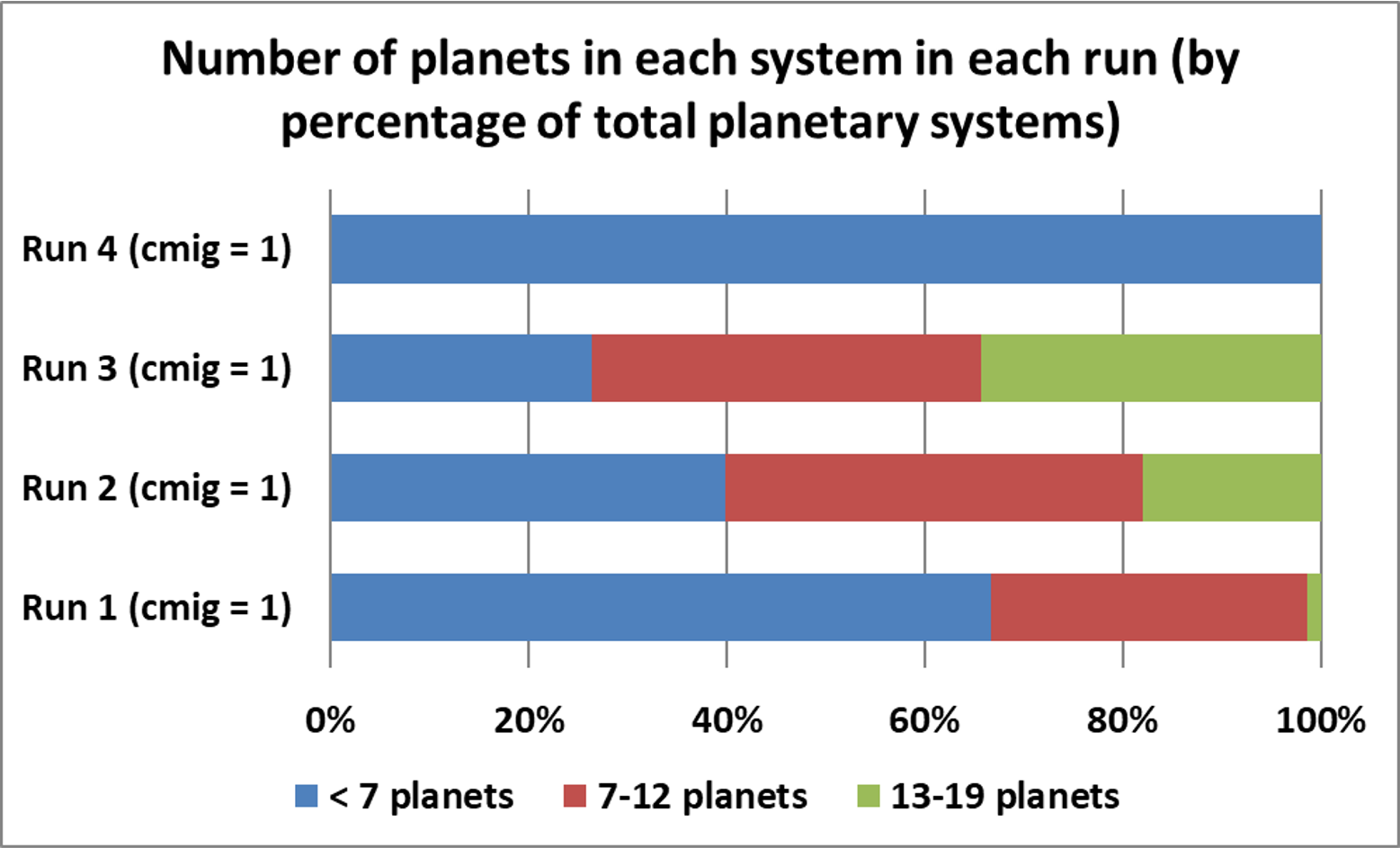}
    \includegraphics[width=\columnwidth]{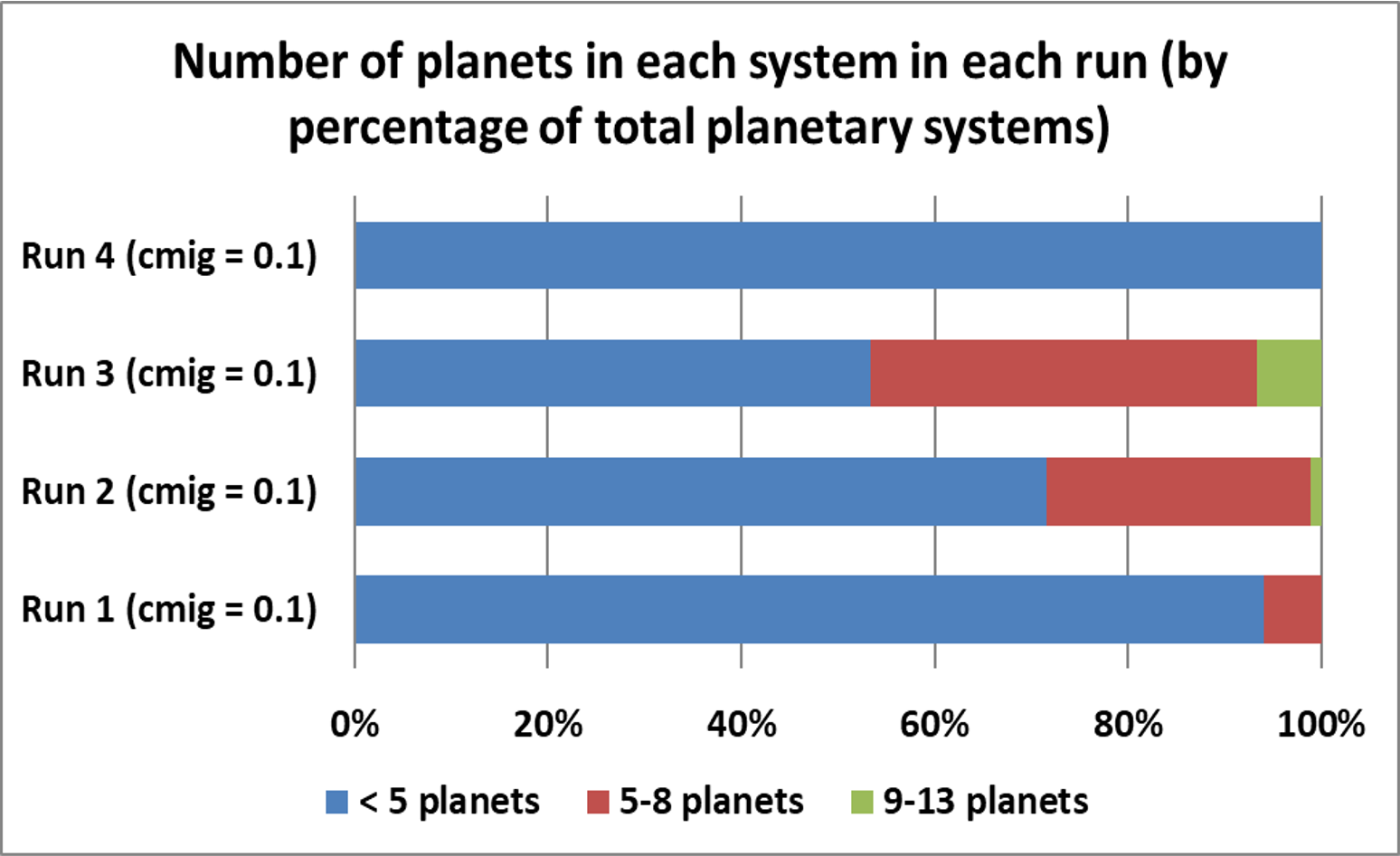}
    \caption{Planetary System architectures by percentage of total planetary systems formed for all runs and both migration models.}
    \label{fig:3}
\end{figure}

To compare our model results with a distribution of observed exoplanets, we utilize a population synthesis approach and run 4 simulations (labeled in Table \ref{tab:1} with $c_{mig}=1$) to construct 1000 planetary systems each using the 4 different disk mass distributions we formulated (Section 2.3). The simulations are shown in Figure \ref{fig:2} along with planets from different observed systems within the stellar mass range assumed. The different types of planetary systems (according to number of planets) are shown in Figure \ref{fig:3}.
\\
\begin{table*}
    \centering
    \begin{tabular}{c|c|c|c}
        \hline
        \hline
       \textbf{Disk mass}  & \textbf{Run} & \textbf{Number of planets} & \textbf{Mean planet mass}\\
       \hline
        3-5\% star mass (\emph{Ormel}) & Run 1 ($c_{mig}=1$) & 5846 & 0.448 $M_{\oplus}$\\
        3-8\% star mass & Run 2 ($c_{mig}=1$) & 8321 & 0.5553 $M_{\oplus}$\\
        3-10\% star mass & Run 3 ($c_{mig}=1$) & 10135 & 0.620 $M_{\oplus}$\\
        \emph{Lupus} & Run 4 ($c_{mig}=1$) & 1001 & 0.011 $M_{\oplus}$\\
        \hline
        3-5\% star mass (\emph{Ormel}) & Run 1 ($c_{mig}=0.1$) & 2742 & 0.897 $M_{\oplus}$\\
        3-8\% star mass & Run 2 ($c_{mig}=0.1$) & 3779 & 0.980 $M_{\oplus}$\\
        3-10\% star mass & Run 3 ($c_{mig}=0.1$) & 4691 & 1.024 $M_{\oplus}$\\
        \emph{Lupus} & Run 4 ($c_{mig}=0.1$) & 1000 & 0.010 $M_{\oplus}$\\
        \hline
    \end{tabular}
    \caption{Different Simulation Runs and results for $c_{mig}=1$ and $c_{mig}=0.1$.}
    \label{tab:1}
\end{table*}
\\
There were 5846 planets simulated in total in Run 1 (\emph{Ormel disks} or disk mass is 3-5\% star mass) with a mean planet mass of 0.448$M_{\oplus}$. All planetary systems formed are multi-planetary with the majority of the systems (66.7\%) having < 7 planets and 31.9\% having 6-12 planets. The plot for Run 1 shows two peaks for population spread of planets: a much larger clustered population of massive planets from 0.01 AU to 0.03 AU on the left and a less clustered less massive planet population spread out almost uniformly > 0.04 AU. The cluster on the left represents the planets now stuck close to the disk edge forming compact planetary convoys. The indication of 'banded' structures in the distribution is a simulation artefact and is a result of planets having preferred positions as the inner edge of the disk doesn't really vary a lot for the stellar masses we consider for our simulations. The size of this cluster indicates that a majority of the simulated planets are able to migrate close to the disk edge and become close-in systems. The simulated mass and semi-major axis range explains the formation of 7 planets which includes TRAPPIST-1 b, c, d, e and h and the first two planets in the YZ Cet system. The rest of the planets remain out of bounds and could indicate requiring more massive disks or a different formation scenario compared to this model.
\\
\\
For Run 2 (disk mass is 3-8\% star mass), there were 8321 planets simulated with a mean mass of 0.553 $M_{\oplus}$. The increase in mean mass is expected as the average disk mass in this run is larger. All planetary systems formed are multi-planetary with the majority of the systems (42.2\%) having 7-12 planets and a close percentage (39.9\%) having < 7 planets. The clustered massive planet distribution on the left (from Run 1) has become even denser and a bit wider as the number of planets in convoys has gone up and is now continuous with the previous less clustered population. The entire TRAPPIST-1, YZ Cet and Teegarden's star systems can now be formed. In addition, Proxima Cen b and Ross 128b can now also be formed.
\\
\\
For Run 3 (disk mass is 3-10\% star mass), 10135 planets were simulated and the mean mass is the highest at 0.620 $M_{\oplus}$. All planetary systems formed are multi-planetary with the majority of the systems (39.4\%) having 7-12 planets but a much increased percentage (34.3\%) of systems, in comparison to all other Runs, also having 13-19 planets. The population of massive planets close to the disk edge is the highest and the cluster now extends the farthest among all runs as the number of planets in convoys is the highest. Even with this increase, the number of planets that can be formed remains the same as in Run 2. All other planets remain unexplained by our model as they are either more massive than the maximum possible upper mass constraint or are just too far away.
\\
\\
For Run 4 (\emph{Lupus disks}), all simulated systems mostly have only 1 planet with only 1001 planets being simulated in total (only one system has 2 planets). The mean simulated planet mass is 0.011 $M_{\oplus}$ which is less than the Mars mass threshold. From Section 3.1.1, this means that most simulated planets are not able to accrete enough mass to enable them to cross the ice-line (seen from the plot by the distribution of planets in between the ice-line width) and no more planets can be formed in a system for the vast majority of the cases. Hence, this disk mass distribution, that is the one taken from observations, doesn't explain the formation of any of the observed planetary systems.

\subsubsection{$c_{mig} = 0.1$(reduced rate migration model)}
\begin{figure*}
    \centering
    \includegraphics[width=\columnwidth]{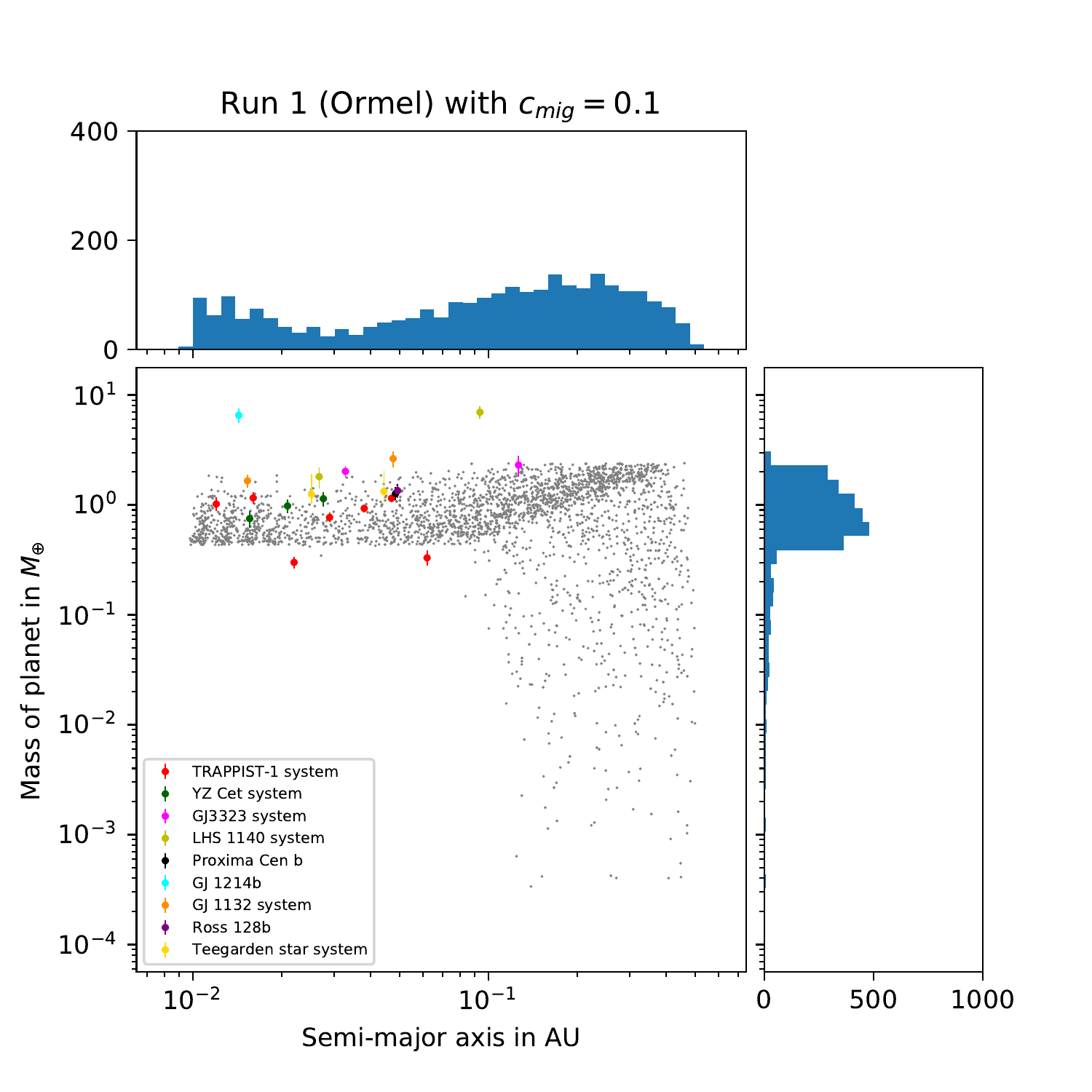}
    \includegraphics[width=\columnwidth]{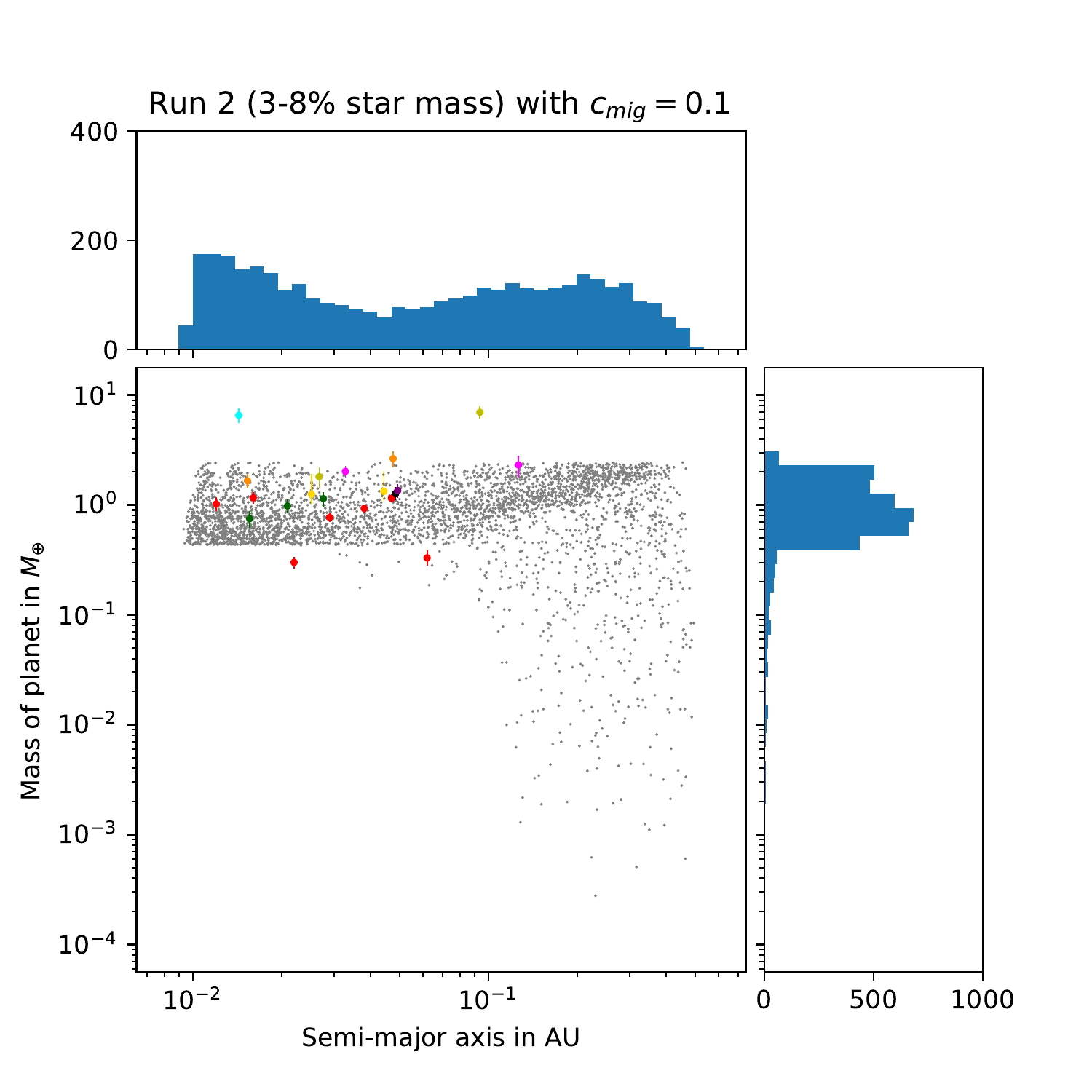}
    \includegraphics[width=\columnwidth]{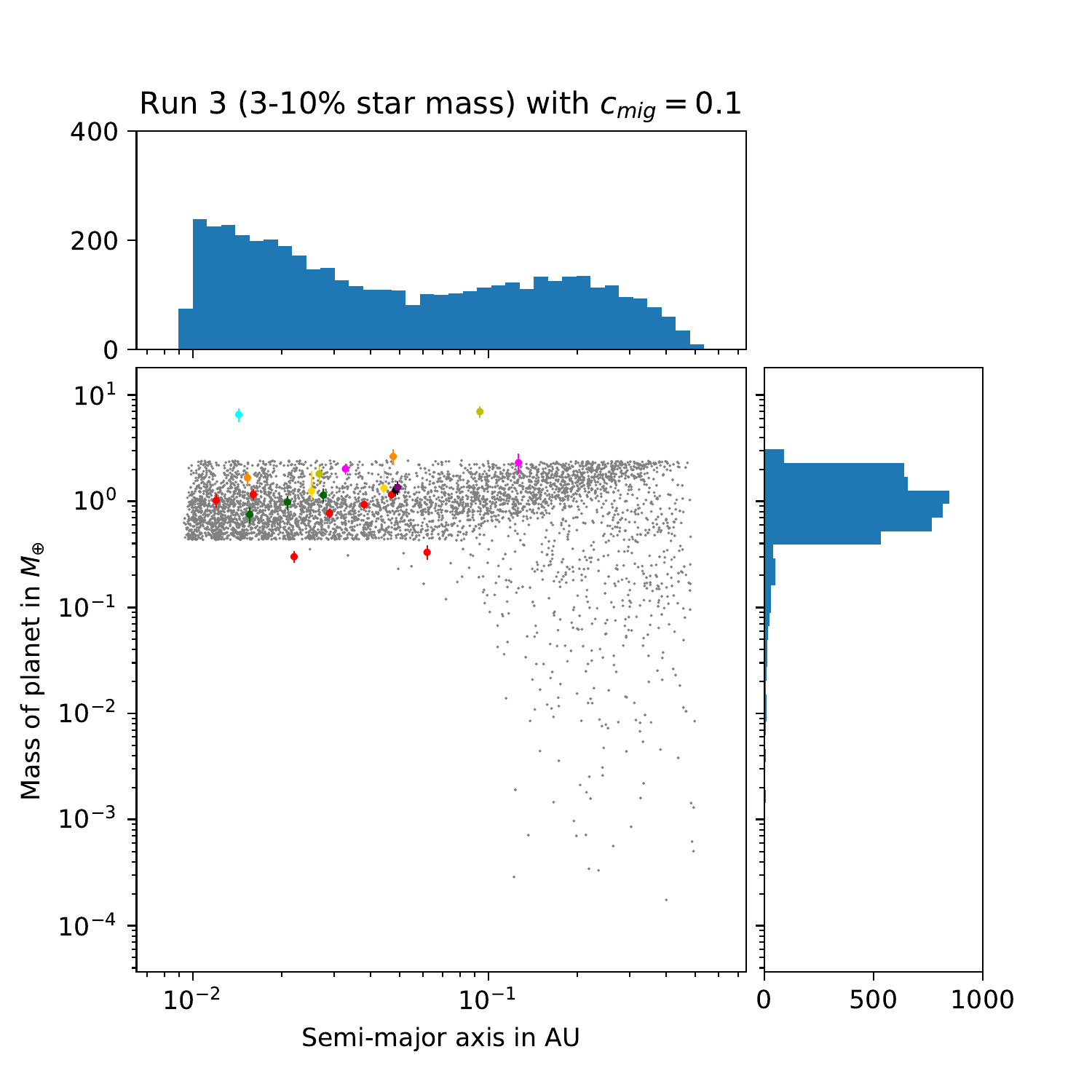}
    \includegraphics[width=\columnwidth]{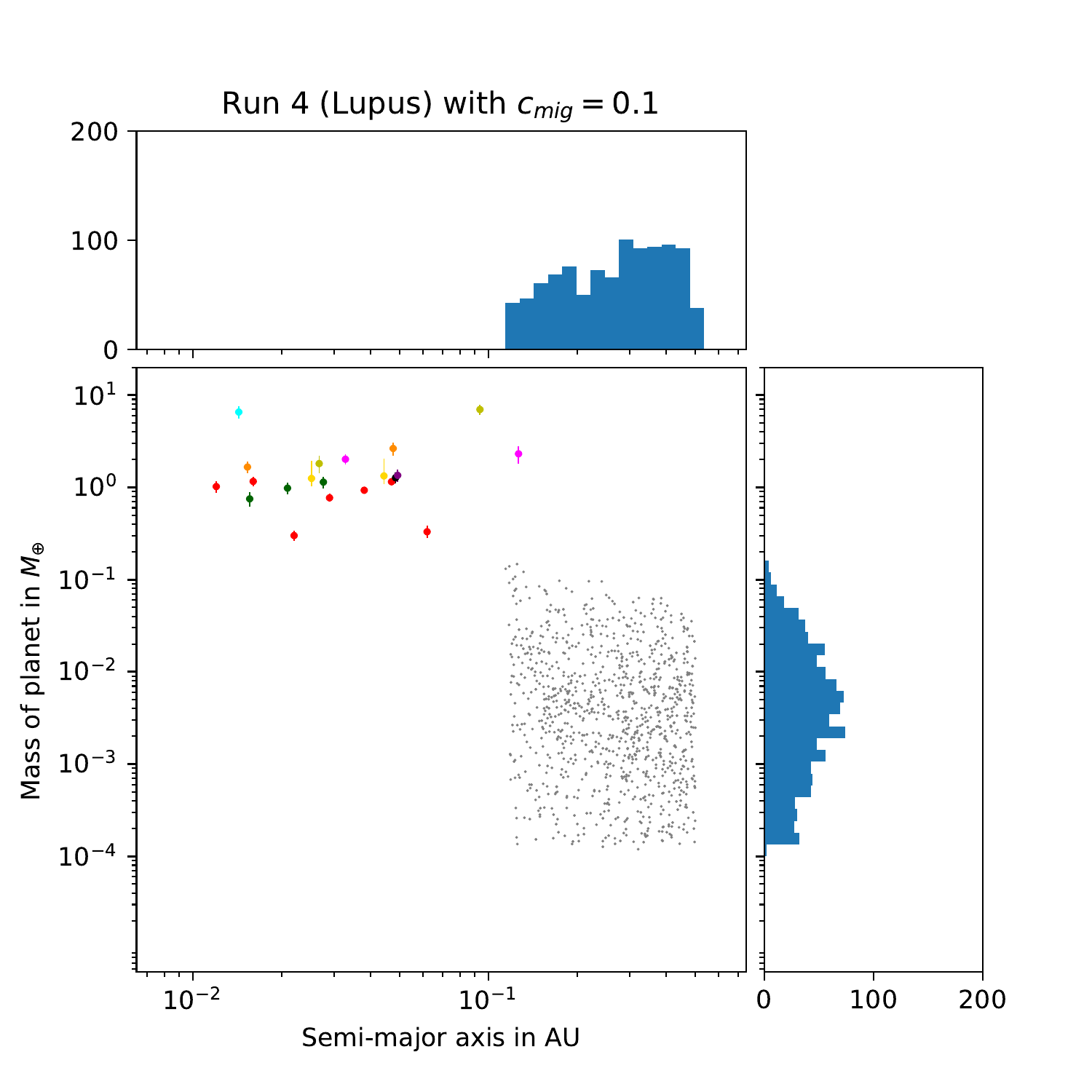}
    \caption{Synthetic population of planets for (top left) \emph{Ormel} Disks (3-5\% star mass), (top right) a bit more heavier disks (3-8\% star mass), (bottom left) a lot more massive disks (3-10\% star mass) and (bottom right) \emph{Lupus} disks all with the migration rate reduced 10 times. The parameters of planets in each observed system are taken from Figure \ref{fig:2}.}
    \label{fig:2b}
\end{figure*}

We run all four simulations again using our delayed migration model, where we delay the migration by a factor of 10. The simulations are shown in Figure \ref{fig:2b} and the different types of planetary systems (according to number of planets) are shown in Figure \ref{fig:3}.
\\
\\
2742 planets were simulated in Run 1 (\emph{Ormel disks} or disk mass is 3-5\% star mass) with a mean planet mass of 0.897 $M_{\oplus}$. This mean mass is higher because most planets reach their \emph{Isolation masses} at higher values of semi-major axes due to slower migration but same accretion rate as compared to its faster migration counterpart (See Section 3.1.2). The slower migration rate is also responsible for reduction in the total number of simulated planets. The planetary systems formed are both single and multi planetary with 94\% of the systems having < 5 planets. The rest of the systems have between 5-8 planets. The distribution of planets is bi-modal with a wide peak from 0.01 - 0.03 AU and a broader peak at around 0.1 AU which is a result of less planets being able to migrate to the disk edge and being stuck close to the ice-line. The plot for Run 1 shows that the number of planets that can now be explained is more than even Run 2 and 3 for the faster/standard migration model with all planets except GJ 1214b, LHS 1140b and c, GJ 3323b, GJ 1132c and the two lower TRAPPIST-1 planets (d and h) now being able to be formed by our model.
\\
\\
In Run 2 (disk mass is 3-8\% star mass), 3779 planets were simulated with a mean planet mass of 0.980 $M_{\oplus}$. Planetary systems formed are both single an multi planetary with 71.5\% of systems having < 5 planets and 27.4\% of systems having between 5-8 planets. The distribution of planets is still bi-modal but the proportion of planets stuck at the disk edge in convoys has increased. In addition to all the planets in Run 1 for the delayed migration model, GJ 3323b, LHS 1140b and GJ 1132c can now also be formed. In addition, the two lowest mass planets of the TRAPPIST-1 system (d and h) now lie in areas with a few simulated planets which shows that they have a low probability of being formed by this model.
\\
\\
In Run 3 (disk mass is 3-10\% star mass), 4691 planets were simulated with a mean planet mass of 1.024 $M_{\oplus}$. Planetary systems formed are both single an multi planetary with 53.4\% of systems having < 5 planets, 39.9\% of systems having between 5-8 planets and 7.4\% having 8-12 planets. The distribution of planets remains bi-modal with the proportion of planets stuck at the disk edge being the highest of all the runs for the reduced rate migration model. The number of planets that can be formed remains the same as Run 2. The two lowest mass planets of the TRAPPIST-1 system (d and h) still lie in areas with a few simulated planets which shows that either lower massive disks or slightly faster migration models are needed to explain their formation. GJ 1214b and LHS 1140c remain out of bounds since they are more massive than the highest massive planet that can be formed due to the \emph{Isolation mass} constraint. In general, this reduced rate migration model accounts for the formation of the maximum number of observed exoplanets (in Run 2 and 3 both).
\\
\\
Run 4 (\emph{Lupus disks}) for the delayed migration model has very similar behaviour as its faster migration counterpart with the mean planet mass and the total number of simulated planets remaining almost the same (0.010 $M_{\oplus}$ and 1000 planets respectively). As discussed in Section 3.1.2 (for the case of \emph{Lupus} disks), this is because mass accretion is not affected by a delayed migration model but planets do not grow beyond the Mars mass limit in both cases and hence there is no substantial migration inwards to affect the mass distribution. So the growth of planets to similar masses at similar locations remains unchanged. This means that even this model is unable to form exoplanets with disk masses from observations.
\section{Discussion}
\subsection{Observed disks not massive enough?}
Section 3.2 and Figures \ref{fig:2} and \ref{fig:2b} show that only Runs 1 through 3 for both the standard and reduced rate migration models can form planets with masses larger than Mars and multiple planets in planetary systems for all cases. This is because of the increased masses of the disk distributions. It can already be observed from Figure \ref{fig:0} that \emph{Lupus disks} are less massive than the rest three distributions. This shows that the more observationally accurate disk masses are simply not high enough for planets to reach the Mars mass threshold and the planet migration is consequently not fast enough for these planets to migrate out of the ice-line. This also means that multiple planet formation is hindered since our model is a sequential model of planetary system formation. The lower limit of the more massive and the upper limit \emph{Lupus} disk mass distributions indicates that the minimum disk mass needed to grow a Mars mass planet is about $2 \times 10^{-3}$ $M_{\odot}$.
\\
\\
Figures \ref{fig:2} and \ref{fig:2b} also provide an interesting observation: Much heavier e.g. $\geqslant$ \emph{Ormel disks} are able to form observed exoplanets while the more precise \emph{Lupus} and less massive disks aren't able to explain any of them. This poses the question of whether there is a significant underestimation of disk masses around low mass stars by sub-mm disk surveys. \citet{manara2018protoplanetary} also came to this conclusion by comparing known exoplanet masses around low mass stars and masses of protoplanetary disks around low mass stars of age 1-3 million years. Assuming that sub-mm emission in disks is optically thin, they hypothesized that either growth to planetesimal and planet sizes occurs rapidly within 1 million years in disks or there is efficient and continuous (or periodic) accretion of material from the environment to the disk which replenishes the material accreted into the star. \citet{zhu2019one} however presented a possibility of scattering in disks being a factor in making optically thick disks appear optically thin which would underestimate mass of disks. Nevertheless, our model in this work shows that with our currently observed disk masses, none of the currently observed exoplanets can be formed. Hence, this remains an open question.
\\
\\
Signatures of dust growth to mm-cm size within a million years in envelopes and inner disk in Class I YSOs have been found by \citet{miotello2014grain} and \citet{harsono2018evidence}. This raises the possibility of \emph{Pebble Accretion} having time to act within this lifetime if planetary seeds to efficiently accrete pebbles are formed by various mechanisms. Based on this, sub-million year planet formation models based on pebble accretion alone or a hybrid pebble-planetesimal accretion mechanism have been proposed for forming super earths and giant planet cores by \citet{voelkel2020impact}, \citet{brugger2020pebbles}, \citet{coleman2019pebbles} and \citet{liu2020pebbledriven}. All this further strengthens the possibility that planet formation in protoplanetary disks starts well within a million years in which case our model predicts that massive disks with mass $\geqslant$ $2 \times 10^{-3}$ $M_{\odot}$ would have to be present within a million years and substantial amount of planet formation would have to occur to get the observed \emph{Lupus} disk masses.
\\
\\
\citet{liu2020pebbledriven} also present the case of massive disks being present very early in the disk lifetime (with a linear \emph{Ormel} disk to star mass relation) by integrating observed gas accretion over the disk lifetime. A steepening of the disk mass to star mass relation has been observed in sub-mm disk surveys of low mass disks by \citet{ansdell2016alma} by comparison between younger (1-3 million years) \emph{Lupus} and \emph{Taurus} disks with the older \emph{Scorpius} disks. Both of these things considered together might indicate that rapid planet formation and growth within a million years might also substantially deplete disk masses even before 1 million years. But a similar steepening of disk-mass to star mass relation between very young (< 1 million years) and young (1-3 million years) low mass disks has not yet been observed. This study motivates the need for more observations of very young low mass disks in the future.
\\
\\
There is always a concern that massive disks around low mass stars early in their lifetime could become gravitationally unstable. However, \citet{haworth2020massive} recently showed from SPH simulations that massive disks with disk to star mass ratios much greater than 0.1 could be stable around such stars even at a young age of 0.5 million years. This bodes well for our model since almost all planet growth occurs within this time frame.

\subsection{Comparison with other models}
We now compare our results with other population synthesis calculations. \citet{miguel2020diverse} assumed a planetesimal accretion based formation model and found a bi-modal distribution of simulated exoplanets in compact systems similar to TRAPPIST-1 around low mass stars and brown dwarfs. This trend is almost absent in our distribution for heavy disks for the standard migration model with a small presence in less massive disks but the reduced rate migration plots have a prominent bi-modal distribution (except \emph{Lupus} disks for both cases). They also found that planets above 0.1 $M_{\oplus}$ (Mars mass) were only possible in disks with masses greater than 0.01 $M_{\odot}$. In comparison, our model predicts that even disks with masses as low as $2\times10^{-3} M_{\odot}$ can form planets greater than 0.1 $M_{\oplus}$. This clearly shows the higher efficiency of the \emph{Pebble Accretion} mechanism. However, the highest possible planet mass in their model is much higher than our model. This is possible due to merger of several earth mass planetesimals. However, we do not take such mergers into account for our model.
\\
\\
Their model as well as our model require a reduction factor of 10 for Type I migration to explain most observed exoplanets. Both of these models also predict an underestimation of disk masses due to an inability to explain all observed exoplanet masses with currently observed disk masses.
\\
\\
\citet{alibert2017formation} also developed a core accretion model and found that compact planetary systems similar to the TRAPPIST-1 system were possible around low mass stars. However, the number of planets in their simulated planetary systems were limited to 4 planets. In comparison, even our reduced rate migration model simulates around 6 planets in the case of \emph{Ormel} disks and 13 planets with even higher massive disks within a 3 million year disk lifetime (for a standard migration model, the number for \emph{Ormel} is 15 and 19 for the more massive disks). This might be due to us taking the starting time of planet formation at t=0. With a larger starting time, we expect the number of planets in our simulated systems to fall. They also found that the disk to star mass correlation and the disk lifetime are important constraints for the kind of planets formed and had a huge impact on the water content of such planets. We also found that the disk to star mass correlation is important for the characteristics of the simulated planet distribution but the disk lifetime doesn't seem to make a difference even for low mass disks as \emph{Pebble Accretion} stops long before the disk lifetime in general. However, we make no comments about the water content of our planet distribution but we expect some amount of water from the predominantly accreted icy pebbles during growth within the ice-line.
\\
\\
\citet{coleman2019pebbles} presented a Pebble accretion driven planet growth model to compare against their own planetesimal accretion based model for forming TRAPPIST-1 analogs around low mass stars and found that these mostly compare favourably with each other. Although they did not formulate a population synthesis approach, the differences they have with our model is the consideration of planet collisions resulting in mergers, taking smaller disks ($a_{out}$ of 10 AU), different locations over the entire inner disk for the initial planet embryo, a much larger embryo mass of $10^{-2} M_{\oplus}$ and a much longer disk lifetime of 10 million years. We have considered much larger disks ($a_{out}$ of 200 AU), no planet collisions or mergers, embryo growth only inside the ice-line width, an initial embryo mass of $10^{-4} M_{\oplus}$ and much shorter disk lifetimes between 1-3 million years. Consequently, our simulated planets have smaller masses due to the \emph{Isolation} mass constraint as no mergers of planetesimals can happen. \citet{coleman2019pebbles} also used more precise N-body simulations to determine the orbits of their non-sequential planet formation model, while we constrain our planets to have orbits determined by the minimum spacing between planets for stable orbits (from empirical evidence) as well as disk dispersal which stops all migration.
\\
\\
More recently, \citet{liu2020pebbledriven} proposed a population synthesis model using \emph{Pebble Accretion} around very low mass stars and brown dwarfs (0.01 to 0.1 $M_{\odot}$) and found mass range of simulated planets to be around an earth mass around 0.1 $M_{\odot}$ stars. This is very similar to our simulated exoplanets by a \textcolor{blue}{reduced rate} migration model around a similar mass star. The planet masses around a 0.1 $M_{\odot}$ star for our standard migration model is lower than this value  which might be a result of the different \emph{Isolation mass} equations we use. Our model is also close to the Scenario A presented in their work where embryo growth starts at the ice-line for self-gravitating disks but with a much higher stellar mass range. Consequently, a majority of our planets grow beyond the 0.1 $M_{\oplus}$ limit and also form a larger population of close-in planets near the disk edge. While their work is focused on analyzing the final mass, semi-major axis and composition of their simulated planets by varying the stellar host masses (and considering only a higher disk mass and much smaller disk radius), we examine the effect of varying disk mass distributions on the simulated final planet mass and semi-major axis distribution and compare it with observed exoplanets with constraints from observations of protoplanetary disks.
\section{Conclusions}
Motivated by the apparent disparity between observed disk masses and masses needed to form exoplanets around low mass stars presented in \citet{manara2018protoplanetary}, we expand an analytical model of planet evolution using the efficient \emph{Pebble Accretion} process for planet mass growth and Type-I migration postulated by \citet{ormel2017formation} and develop it into a population synthesis model. We find that compact resonant multi-planetary systems like TRAPPIST-1 will be common around low mass stars within a 3 million year old disk lifetime if the disk masses are higher than what is being currently observed by sub-mm disk surveys of low mass disks. In addition, migration delayed by a factor of 10 for heavier disks is able to account for most observed exoplanets around low mass stars.
\\
\\
The type of distribution of disk masses also influences the simulated planet distribution with heavier disks accounting for more observed planets in general. The minimum mass required to grow planets beyond a Mars mass threshold and form multi-planetary systems is $2\times10^{-3}$ $M_{\odot}$ which is two times higher than the gas disk mass range for observed protoplanetary disks from sub-mm surveys around low mass stars (from Figure \ref{fig:i} top panel). This points to either an underestimation of mass in disks by sub-mm disk surveys or a rapid planet formation and growth process that depletes the mass in disks even before 1 million years which while being an integral component of several pebble accretion and hybrid pebble-planetesimal accretion based planet formation models, cannot still be accounted by these disk surveys as of now. This motivates the observation of more embedded disks in low mass star forming regions in systems younger than a million years.
\\
\\
We also compare our results to other population synthesis models based on planetesimal accretion mechanism and find that our model can form more planets in multi-planetary systems. Comparison with other pebble accretion models shows that the upper mass range and close-in nature of our simulated planets are in line with most of these models unless planet mergers or different \emph{Isolation mass} equations are taken into account. In future, we expect mergers to be accounted for and more computationally intensive N-body simulations to be used for planet orbit distribution using similar population synthesis models.
\section*{Data Availability}
All Python codes used to generate simulated planet data used in this paper are freely available online in a github repository linked in the Methods section (See footnote for Section 2.3). Disk mass data used for generating Figure \ref{fig:i} are available online as part of supplementary material for the \citet{ansdell2016alma} paper.




\bibliographystyle{mnras}
\bibliography{references} 








\bsp	
\label{lastpage}
\end{document}